\documentclass[12pt]{article}

\usepackage{newtxtext,newtxmath}
\usepackage{booktabs}
\usepackage{graphicx}

\usepackage[letterpaper,margin=1in]{geometry}

\renewenvironment{abstract}
	{\quotation}
	{\endquotation}

\date{}

\makeatletter
\renewcommand{\fnum@figure}{\textbf{Figure \thefigure}}
\renewcommand{\fnum@table}{\textbf{Table \thetable}}
\makeatother

\usepackage{scicite}

\usepackage{url}
\usepackage{lineno}

\def\scititle{
	Reconstructing Historical Climate Fields With Deep Learning
}
\title{\bfseries \boldmath \scititle}

\author{
	Nils~Bochow$^{1,2,3\ast}$,
	Anna~Poltronieri$^{1}$,
	Martin~Rypdal$^{1}$,
    Niklas~Boers$^{3,4,5}$\and
	\small$^{1}$Department of Mathematics and Statistics, Faculty of Science and Technology, \and \small UiT - The Arctic University of Norway, Tromsø, Norway.\and
	\small$^{2}$Physics of Ice, Climate and Earth, Niels Bohr Institute, University of Copenhagen, Copenhagen, Denmark.\and
 	\small$^{3}$Potsdam Institute for Climate Impact Research, Potsdam, Germany.\and
    \small$^{4}$Earth System Modelling, School of Engineering, Design, Technical University of Munich,  Munich, Germany.\and
    \small$^{5}$Department of Mathematics and Global Systems Institute, University of Exeter, Exeter, UK.\and
	\small$^\ast$Corresponding author. Email: contact@nilsbochow.com
}

\begin{document} 

\maketitle

\begin{abstract} \bfseries \boldmath
Historical records of climate fields are often sparse due to missing measurements, especially before the introduction of large-scale satellite missions. Several statistical and model-based methods have been introduced to fill gaps and reconstruct historical records. 
Here, we employ a recently introduced deep-learning approach based on Fourier convolutions, trained on numerical climate model output, to reconstruct historical climate fields. 
Using this approach, we are able to realistically reconstruct large and irregular areas of missing data, and to reproduce known historical events, such as strong El Niño or La Niña events, with very little given information. 
Our method outperforms the widely used statistical kriging method, as well as other recent machine learning approaches. 
The model generalizes to higher resolutions than the ones it was trained on and can be used on a variety of climate fields. 
Moreover, it allows inpainting of masks never seen before during the model training.
\end{abstract}

\section*{Introduction}
Observational climate data is typically sparse before systematic observations such as buoys, ship measurements, or satellite measurements were introduced. 
Generally, the further back in time we go, the fewer observations are available \cite{ben-yami_uncertainties_2023}. 
Temperature and precipitation records are the best-observed climate fields in the recent past and reach back until the 19th century, but measurements are still sparse and rely heavily on interpolation, especially for earlier parts of the records \cite{morice_quantifying_2012, harris_version_2020}.
Even more severely, for many important climate variables, such as sea-ice thickness or vegetation indices, no measurements exist at all before the introduction of large-scale satellite missions. 
The corresponding time series often span a few decades or even only years \cite{vermote_noaa_2018, landy_year-round_2022}. 
The low spatial and temporal resolution introduces large uncertainties and limits our understanding of important climatic processes \cite{morice_quantifying_2012, cowtan_coverage_2014, ben-yami_uncertainties_2023}.

Several approaches and methods to produce historical climate fields based on the available observations have been developed in the past. 
One approach is to run state-of-the-art weather models with observations and past weather forecasts to produce reanalysis products that provide a complete picture of the past weather and climate for the last decades \cite{kalnay_ncepncar_1996, compo_twentieth_2011, bell_era5_2021, soci_era5_2024}.
While reanalyses are successful in providing spatiotemporally continuous and consistent data, they often struggle with specific regions and variables and inherit biases the employed numerical models suffer from \cite{bell_era5_2021, slivinski_evaluation_2021}.

An alternative approach is to use statistical methods to reconstruct missing information. 
In this regard, kriging or Gaussian process regression is widely used in the geosciences \cite{berezowski_cplfd-gdpt5_2016, sekulic_spatio-temporal_2020, belkhiri_spatial_2020}.
However, statistical methods typically do not include knowledge of the temporal and spatial patterns of the underlying climatic fields and therefore fail to reconstruct these patterns, especially for large missing areas. 

In recent years, machine learning (ML) has become widely used in geoscience and climate science, with the promise of better performance than statistical methods while still providing easy usability and, to some extent, knowledge of the underlying physical processes \cite{kadow_artificial_2020,irrgang_towards_2021}.
The applications of ML in climate science are vast and range from classical time series forecasting \cite{mitsui_seasonal_2021,lim_time-series_2021, lam_learning_2023}, down-scaling and post-processing of numerical models \cite{grattarola_generalised_2022, hess_physically_2022}, to time series reconstruction \cite{huang_reconstructing_2020, kadow_artificial_2020}.
Furthermore, there is a substantial ongoing effort to combine traditional numerical Earth system models with ML methods to leverage the advantages of both approaches \cite{monteleoni_climate_2013, reichstein_deep_2019, irrgang_towards_2021, yuval_use_2021, zhu_physics-informed_2022, gelbrecht_differentiable_2023, schneider_harnessing_2023, de_burgh-day_machine_2023, kochkov_neural_2023}. 

In this study, we consider the reconstruction of spatial climate fields as an image inpainting problem.   
Inpainting images based on given information is a classical problem in computer vision and many approaches have been proposed in recent years \cite{qin_image_2021, zhang_-gan_2022, chen_rnon_2023}. 
We apply the recently introduced state-of-the-art deep learning approach \textit{Resolution-robust Large Mask Inpainting with Fourier Convolutions} (LaMa) \cite{suvorov_resolution-robust_2021} to reconstruct different climate fields with a focus on surface temperature records. 
We train our model on numerical climate model output from the Coupled Model Intercomparison Project to reconstruct the missing measurements in observational data.
Our method is able to reconstruct climate fields with very sparse information and highly irregular missing data. 
We show that our approach outperforms kriging and other ML methods. Moreover, it is able to inpaint different data sets than the ones it was trained on, and can be used on a variety of structurally different climatic fields at varying resolutions. 

\begin{figure}[t]
\includegraphics[width=\textwidth]{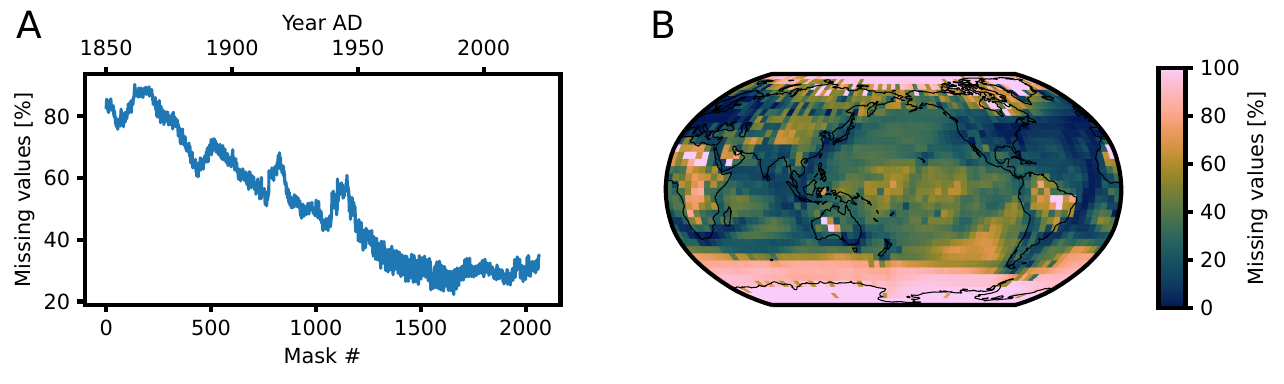}
\caption{\textbf{Spatial map and temporal time series of missing observations in HadCRUT4.} \textbf{(A)} Time series of the missing value ratio on the grid cell level in HadCRUT4 for the whole Earth. There is a steady increase in the observational temperature coverage with some exceptions such as the two world wars. \textbf{(B)} Spatial missing ratio in HadCRUT4 over the whole time span from 1850-2022. The polar regions show the lowest coverage of temperature records.}
\label{fig:missing_hadcrut}
\end{figure}

The surface temperature is one of the most important climate variables, as a direct measure of climate change. 
Global instrumental temperature records reach back to the mid-19th century \cite{morice_quantifying_2012} with local observations reaching back as far as the mid-17th century \cite{parker_new_1992}.
However, on average, less than 30\% of Earth's surface before the year 1900~AD have measurements in the state-of-the-art observational data set HadCRUT4 (Fig.~\ref{fig:missing_hadcrut}A). 
This is similar for other widely used long-term temperature data. 
Therefore, surface temperature records serve as perfect proof-of-concept applications for the image inpainting task in climate science. 

\section*{Results}
To inpaint the temperature records, we train our model on the historical surface temperatures from the Coupled Model Intercomparison Project 5 (CMIP5) ensemble (1850–2012~AD); see Methods. 
We follow a previously introduced mask generation approach \cite{kadow_artificial_2020} and mask the training data with the missing masks derived from the observational gaps in the temperature data set HadCRUT4 \cite{morice_quantifying_2012} during training. 
 
First, we evaluate the model on the same held-out CMIP5 member as in \cite{kadow_artificial_2020} to directly compare with their inpainting approach, which is based on partial convolutions (PConv) and trained on the same CMIP5 training set.
In a second step, we evaluate the trained model on each HadCRUT4 mask for 2251 randomly held-out months of the CMIP5 ensemble. 
This gives a total of 4,641,750 combinations of images and masks.
As a baseline comparison, we compare our results with statistical kriging.
Subsequently, we reconstruct the HadCRUT4 data and show examples of other applications. 

\subsection*{Comparison with related work}
\begin{figure}[t]
\includegraphics[width=\textwidth]{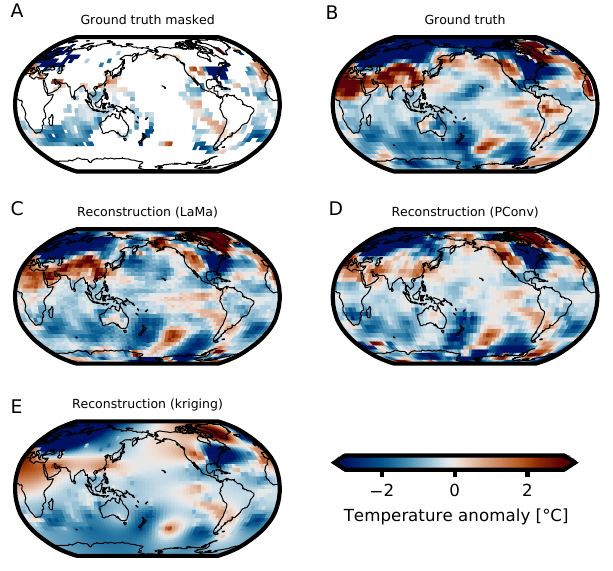}
\caption{\textbf{Example reconstructed CMIP5 temperatures for all methods.} \textbf{(A)} Masked ground truth for February 1870 derived from a held-out CMIP5 member, masked with the corresponding HadCRUT4 mask for this date. White areas denote masked regions. \textbf{(B)} Ground truth without masking. \textbf{(C)} Infilled temperatures via LaMa. The spatial patterns are very similar to the ground truth. \textbf{(D,E)} Same as \textbf{(C)} but for PConv \cite{kadow_artificial_2020} and kriging, respectively. While the spatial patterns reconstructed by the deep learning methods are very similar to the ground truth, some regions show the opposite trend in the temperature e.g., northern South America.}
\label{fig:Fig_2}
\end{figure}

In order for our method to be directly comparable to the previously introduced PConv method \cite{kadow_artificial_2020}, we evaluate LaMa on the same held-out CMIP5 ensemble member.
We mask the held-out 145th CMIP5 member with the corresponding HadCRUT4 masks for each month over the time span 1870-2005~AD, in order to have the same temporal range as Kadow et al. \cite{kadow_artificial_2020} (Fig.~\ref{fig:Fig_2}A,B). 
We reconstruct the held-out CMIP5 member over the whole time span with LaMa and kriging, and compare it with the PConv approach (Fig.~\ref{fig:Fig_2}). 
We use the spatial root-mean-squared error (RMSE) and site-wise RMSE as main evaluation metrics, where the main difference between the metrics is the order of averaging in time and space (see Methods).
We refer to the square root of the spatially weighted average of the squared differences between the ground truth and the inpainted image as the spatial RMSE. 
Additionally, we define the site-wise RMSE as the RMSE at each site (i.e., grid cell), averaged over the time dimension. 
The mean site-wise RMSE is the spatially weighted average of the site-wise RMSE in all grid cells. 
In each time step, we exclude the grid cells that have known values for the calculations of the RMSE.  

\begin{figure}[t]
\includegraphics[width=0.5\textwidth]{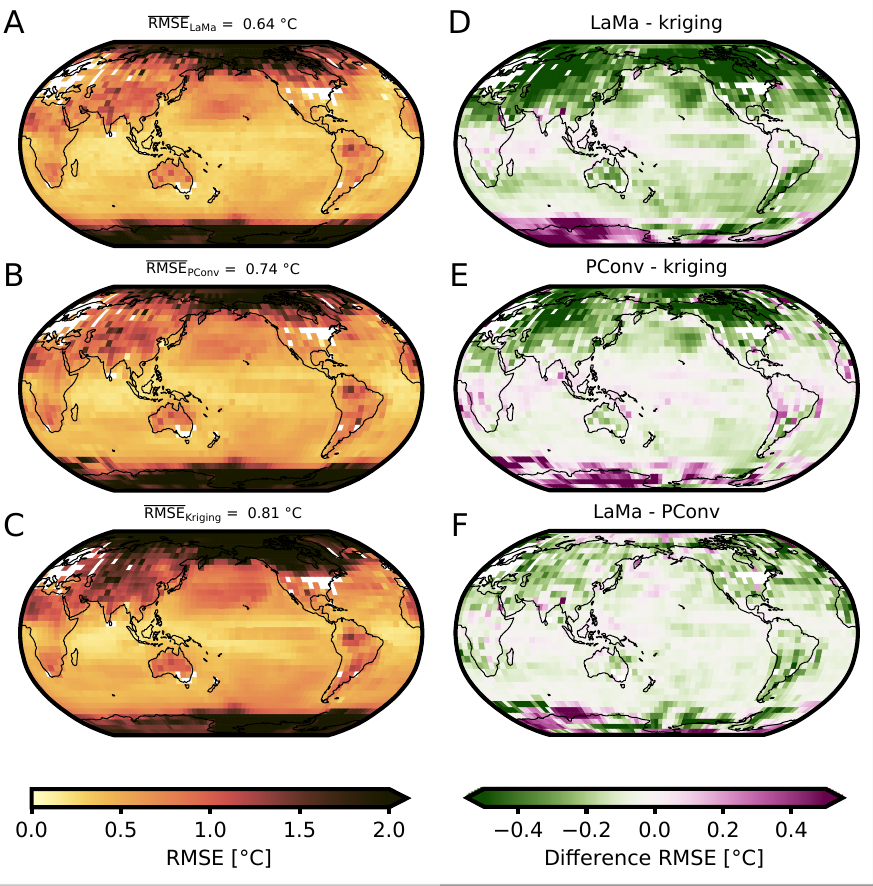}
\caption{\textbf{Average site-wise root-mean-square error (RMSE) and comparison between methods for single held-out CMIP5 member.} \textbf{(A)} Temporally averaged RMSE at each site for inpainted CMIP5 held-out member using LaMa fixed. The white areas denote the regions with available temperature records for the whole period 1870-2005~AD. The spatially weighted average of the site-wise RMSE is $0.64^\circ$C. The polar regions, especially Antarctica, show the greatest RMSE, while the tropical and subtropical regions show the smallest RMSE. \textbf{(B, C)} Same as \textbf{(A)} but for PConv \cite{kadow_artificial_2020} and kriging. Both ML methods show a smaller mean RMSE than kriging. Especially in the Northern Hemisphere the RMSE is smaller. \textbf{(D)} Difference between temporally averaged RMSE at each site for LaMa and kriging. Green areas denote regions where the RMSE of LaMa is smaller than that of the baseline kriging method. Purple areas denote greater RMSE than for kriging. LaMa fixed shows a lower RMSE than kriging in 79\% of the grid cells. \textbf{(E)} Same as \textbf{(D)} but for PConv \cite{kadow_artificial_2020}. PConv shows in 71\% of the grid cells a smaller RMSE than kriging. \textbf{(F)} Comparison between LaMa and PConv. Green areas denote regions where the RMSE of LaMa is smaller than the RMSE of the PConv method. LaMa shows a lower RMSE than PConv in 78\% of all grid cells. }
\label{fig:Fig_3}
\end{figure}

Our model is able to realistically reconstruct the spatial patterns and amplitude of the surface temperature. 
An example spatial reconstruction for February 1870 for all methods, i.e. LaMa, kriging, and PConv \cite{kadow_artificial_2020}, is shown in Fig.~\ref{fig:Fig_2}. 
There is high agreement between the reconstructed temperature fields and the ground truth. 
While the tropical and sub-tropical regions show strong similarities between ground truth and reconstruction, the polar regions show the strongest deviation from the ground truth for all methods (Fig.~\ref{fig:Fig_3}, Tab.~\ref{tab:comparison}).

All ML methods show an improvement compared to kriging in terms of the mean site-wise and mean spatial RMSE (Fig.~\ref{fig:Fig_3}, Tab.~\ref{tab:comparison}). 
LaMa outperforms all other methods with a 15\% lower mean site-wise RMSE than the reference PConv method (Fig.~\ref{fig:Fig_3}F, Tab.~\ref{tab:comparison}). 
LaMa shows the largest improvement in the Northern Hemisphere, especially in North America and Asia, but also in the subtropical Southern Hemisphere and in west Antarctica compared to kriging (green regions in Fig.~\ref{fig:Fig_3}D). 
Note that Eastern Antarctica is in the west in the geographic projection used in these maps here. 
LaMa and the PConv method show a lower site-wise RMSE in 79\% and 71\% of the grid cells compared to kriging, respectively. 
For all methods, the temporally averaged site-wise RMSE is largest in the Arctic and Antarctic region. 
Interestingly, the PConv approach shows a slightly worse performance in the northern and southern tropical Pacific than kriging in terms of the site-wise RMSE (Fig.~\ref{fig:Fig_3}E), while LaMa shows better performance in this region. 
Both ML methods show considerably worse performance in East Antarctica and slightly worse performance in the Indian Ocean than kriging.  
The weaker performance in the aforementioned regions is likely a result of the high-temperature variation in these regions. 
Kriging, by definition, produces a smooth temperature field that might be closer to the ground truth for highly temporally variable regions. 
Especially for Antarctica the reconstructed temperature fields via kriging are very homogeneous due to the distance of known measurements, while the ML approaches inpaint a highly variable Antarctica as learned from the CMIP5 data (Fig.~\ref{fig:Fig_2}). 
This suggests that the available temperature information fed into the networks is not sufficient to successfully infer the temperature patterns in Antarctica.

In 78\% of the grid cells, LaMa shows a lower site-wise RMSE than the reconstruction based on PConv (Fig.~\ref{fig:Fig_3}F).
LaMa is also able to reconstruct the global mean temperature (GMT) time series reasonably and closely follows the ground truth (fig.~\ref{fig:kadow_timeseries_comparison}).
LaMa shows comparable performance to the PConv method \cite{kadow_artificial_2020} with slightly worse RMSE of the yearly GMT, but lower spatial RMSE and higher correlation between the yearly GMT time series than PConv and kriging (fig.~\ref{fig:kadow_timeseries_comparison}). 
The lower spatial RMSE of LaMa's reconstruction implies that LaMa is generally better in reconstructing the spatial temperature patterns. On the other hand, the slightly greater RMSE of the GMT time series implies that LaMa shows slightly worse performance in reconstructing the temporal variability on a global scale. 

\subsection*{Evaluation on CMIP5}

\begin{figure}[t]
\includegraphics[width=\textwidth]{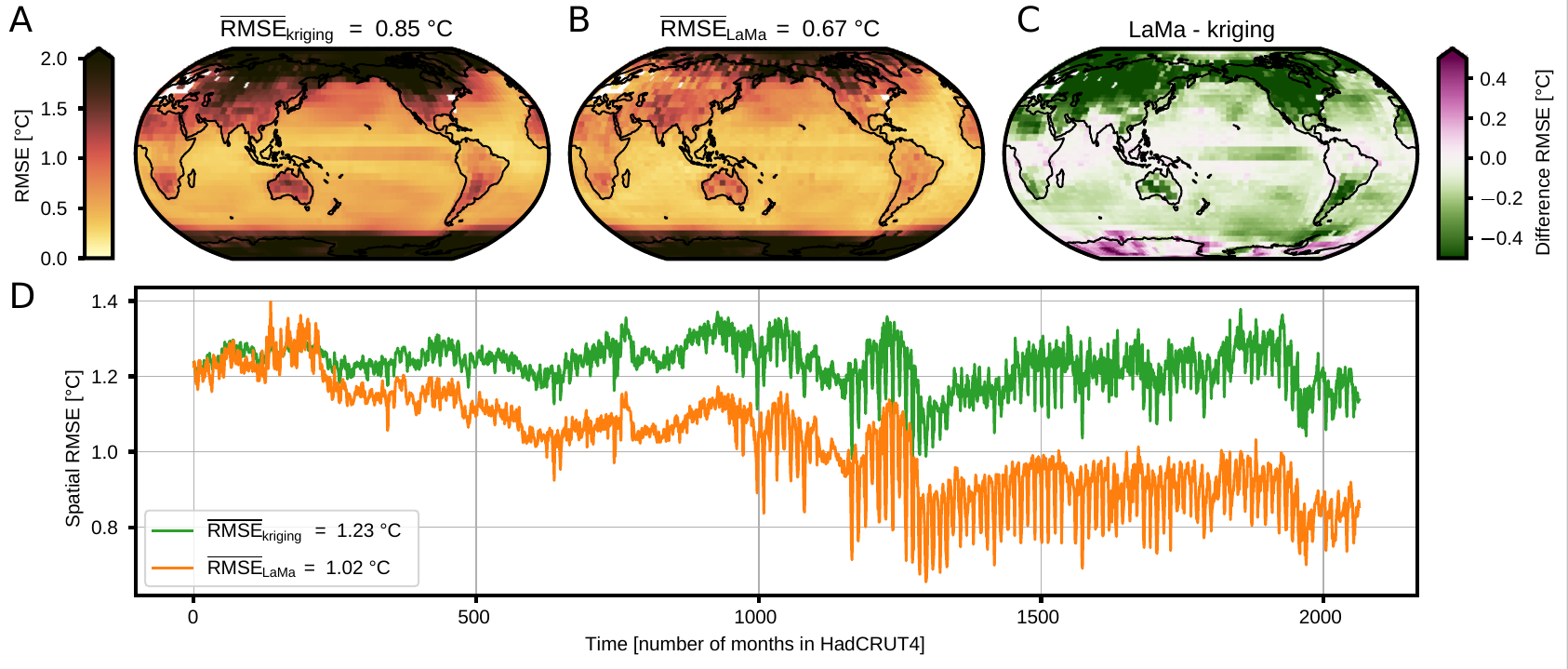}
\caption{\textbf{Error statistics on the held-out CMIP5 members for each HadCRUT4 mask.}\textbf{(A)} Average site-wise RMSE for kriging on the randomly held-out 2251 months from the CMIP5 ensemble. We calculate the site-wise RMSE for every combination of HadCRUT4 masks from 1850 to 2021~AD and month. The spatially weighted average of the site-wise RMSE is $0.85^\circ$C. The polar regions show the largest RMSE, while the tropics and subtropical oceans show the lowest RMSE. White grid cells denote regions where temperature observations are available for the whole time span. \textbf{(B)} Same as \textbf{A} but for LaMa. The site-wise RMSE is lower in most grid cells than for kriging. Especially, in the Northern Hemisphere there is a strong improvement compared to kriging. LaMa outperforms kriging in terms of the spatially averaged site-wise RMSE. \textbf{(C)} Difference between temporally averaged RMSE at each site for LaMa and kriging. Green areas denote regions where the RMSE of LaMa is smaller than that of the baseline kriging method. Purple areas denote greater RMSE than for kriging. \textbf{(D)} Spatial RMSE for both methods and all HadCRUT4 masks, which are ordered in time; note that generally the size of the masks in terms of the number of missing data declines over time. LaMa outperforms kriging almost consistently for all masks. LaMa shows a higher RMSE than kriging for very large masked areas ($\geq80$\%) but outperforms kriging otherwise. Especially for small masks, the average spatial RMSE is substantially lower for LaMa than for kriging.}
\label{fig:Fig_4}
\end{figure}

In addition to the evaluation on the single held-out CMIP5 ensemble member, we evaluate the model on 2251 held-out months of the entire CMIP5 ensemble against all 2064 HadCRUT4 masks. 
We calculate the spatial and site-wise ensemble RMSE of the infilled evaluation temperature data and compare it with kriging (Fig.~\ref{fig:Fig_4}). 
Here, \textit{ensemble mean} refers to the mean over all months of the 2251 CMIP5 ensemble members, and \textit{temporal mean} refers to a mean over the masks, which corresponds to the temporal dimension of the HadCRUT4 records. 

LaMa has a temporal ensemble mean of the spatial RMSE of 1.02~K and the reconstructed images via kriging have a mean spatial RMSE of 1.23~K (Fig.~\ref{fig:Fig_4}D). 
The ensemble mean of the site-wise RMSE shows a similar behavior (Fig.~\ref{fig:Fig_4}A,B,C).
LaMa shows a more than 20\% smaller mean site-wise RMSE than kriging (Fig.~\ref{fig:Fig_4}A,B,C) but shows a higher spatial RMSE than kriging for very large masks ($>80$\%).
Otherwise, LaMa consistently outperforms kriging for all masks.

The spatial RMSE of the infilled images via kriging and LaMa depends on the ratio of missing values and shows a decrease around mask number 1200-1300, which corresponds to the year 1950-1960~AD (Fig.~\ref{fig:Fig_4}D). 
This is due to the introduction of large-scale observational instruments and therefore greater coverage of temperature observations.
Both methods show a seasonal dependency of the spatial RMSE (Fig.~\ref{fig:Fig_4}D) due to a difference in the seasonal global temperature coverage, mostly in the polar regions.
The summer months in the polar regions have a greater coverage than the winter months. 
This leads to a higher spatial RMSE for the austral winter than in the boreal winter when observations in Antarctica are sparse, since Antarctica is the region with the largest uncertainty. 

The spatial patterns of the site-wise RMSE are similar to the site-wise RMSE of the single member test set, with the maximum RMSE close to the poles and a minimum in the tropical and subtropical regions (Fig.~\ref{fig:Fig_4}A,B).
In particular, the RMSE in the Northern Hemisphere is notably smaller for LaMa than for kriging. 

It is important to note that LaMa does generally not have any information about the temporal order of the training set. 
That is, even when randomly permuting the fields of the training set, the results would remain the same. 
However, when randomly selecting months for evaluation from the training and validation set, there might be hidden temporal information in the previous or subsequent months due to the temporal correlation of individual months. 
To rule out this possibility, we first show that the local similarity of consecutive months of the same CMIP5 model is generally low.
We demonstrate this using several similarity metrics, including the correlation between two consecutive months (fig.~\ref{fig:similarity_metrics}B-E).
We also show that when filling in the missing grid cell values of a given month by the previous, following, or the same month of the following year, the RMSE is on average substantially larger than when infilling via LaMa (fig.~\ref{fig:similarity_metrics}A). 
However, this test of local similarity does not consider the known correlation on long spatial scales between adjacent months.
To show that our test set does not contain any temporal information and is thus strictly independent from the training set, we additionally train our model on a different train-validation-test split, where we hold out 17 random years from the training set instead of every 9th month (cf.~Methods).
We repeat our analysis on the 145th held-out CMIP5 member as well as on the alternative training set. 
In both cases, the error metrics are very similar (fig.~\ref{fig:alternative_test_set}).
The model trained on the alternative training-validation-test split partially even outperforms the model shown in the main text. 
These tests together show that our approach for reconstructing the field of a given month does not rely on simply using similar months from the training data. 

\subsection*{HadCRUT4}
After demonstrating that LaMa is able to reconstruct spatial and temporal patterns of the CMIP5 temperatures, we apply the trained network to the HadCRUT4 observational data. 
We show that our method is able to accurately reconstruct the spatial and temporal patterns of the HadCRUT4 data set.
As there is no control data for the reconstructed temperature observations, we first compare with reconstructed HadCRUT4 temperatures via kriging \cite{cowtan_coverage_2014} and also analyze spatial patterns in the reconstruction, focusing on known historical events. 

\begin{figure}[t]
\includegraphics[width=\textwidth]{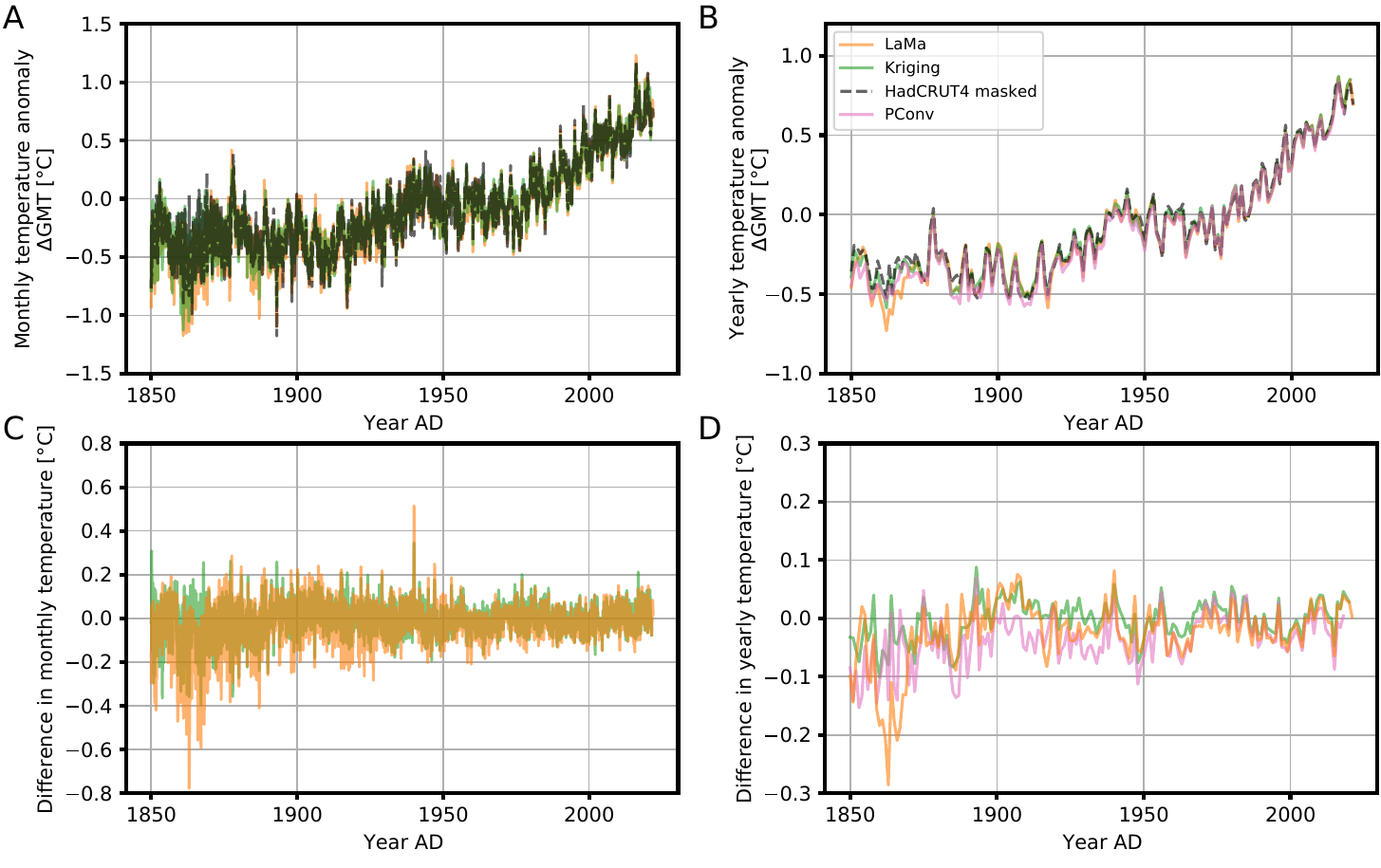}
\caption{\textbf{Reconstructed HadCRUT4 global mean temperature time series.} \textbf{(A)} Monthly HadCRUT4 timeseries from 1850 to 2022 with anomalies relative to the years 1960-1990. The reconstructions from LaMa and kriging as well as the mean of the masked HadCRUT4 records are shown. The dashed black curve is the spatially averaged GMT derived from the incomplete HadCRUT4 observations. \textbf{(B)} Same as \textbf{A} but for the yearly averaged GMT. Additionally, we show the reconstruction based on the PConv method. The Pearson correlation between the masked time series, i.e., yearly GMT and LaMa is $r_\text{LaMa}=0.99$, for kriging $r_\text{kriging}=0.99$ and $r_\text{PConv}=0.99$.
\textbf{(C,D)} Same as \textbf{A,B} but for the difference between the reconstructions and the masked HadCRUT4 time series. }
\label{fig:Fig_5}
\end{figure}

We take a spatially weighted mean to obtain the temporal time series of the GMTs for all methods. 
The reconstructed yearly GMTs show a strong correlation (Pearson correlation coefficient $r>0.99$) and the same trend as the masked HadCRUT4 temperature time series for all methods (Fig.~\ref{fig:Fig_5}).
LaMa shows a lower GMT for the mid-19th century relative to the masked mean and the temperature reconstructed via kriging or PConv. 
The main contributor is a slightly colder Antarctica in LaMa's temperature reconstruction compared to the other methods (fig.~\ref{fig:ts_hemispheres}B,D). 
There is no a priori reason to believe that the global mean time series reconstructed by LaMa is unreasonable. 
However, given another validation on the HadCRUT5 dataset that we carry out (fig.~\ref{fig:hc5_2052_map}\&~\ref{fig:hc5_2052_ts}), and given the underestimation of temperatures in Antarctica on the validation CMIP5 member (Fig.~\ref{fig:Fig_3}), it is likely that the GMT in the mid 19th-century, as reconstructed by LaMa, is underestimated.

Due to the nature of the data, there is no ground truth that we can compare our reconstructions to. 
Hence, we compare the reconstructed temperature fields to well-known historical events such as strong El Niño episodes. 
The El Niño in the year 1877/1878~AD is known to have been extraordinarily strong and is linked to famines around the globe \cite{davis_late_2002, huang_how_2020}. 
However, historical temperature records for these years are sparse.
LaMa is able to reproduce the warm Pacific Ocean based on the sparse records, whereas kriging is not able to reconstruct the spatial patterns of the temperature anomaly (fig.~\ref{fig:november_1877_hadcrut4}).   
The reconstruction based on partial convolutions, PConv (fig.~\ref{fig:november_1877_hadcrut4}E), also shows a warm Pacific but with a smaller spatial extent. 
We also show an opposite example of a strong La Niña year with a cold Pacific for February 1917~AD \cite{giese_nino_2011, voskresenskaya_qualitative_2015} (fig.~\ref{fig:february_1917_hadcrut4}).
Kriging does not reconstruct the same spatial extent of the cold Pacific compared to LaMa, which shows a strongly anomalous cold Pacific (fig.~\ref{fig:february_1917_hadcrut4}). 
Statistical interpolation tends to inpaint large missing areas with values close to zero (e.g., fig.~\ref{fig:november_1877_hadcrut4}).
However, even for these anomalous historical events, it is still hard to verify the validity of the reconstructed temperature anomalies. 
We compare our example reconstructions visually with the 20CRv3.SI reanalysis \cite{compo_twentieth_2011, slivinski_towards_2019} (fig.~\ref{fig:november_1877_hadcrut4}F \& ~\ref{fig:february_1917_hadcrut4}F). 
The ML reconstructions for the two example months show a strong similarity with the reanalysis, while the kriging reconstruction does not show the same spatial patterns. 
This suggests that LaMa is indeed able to capture the dynamics underlying the global temperature fields.
However, the temperature anomalies in Antarctica, in particular, show different spatial patterns across the different reconstructions and temperature products. 
It has to be noted that the 20CRv3.SI reanalysis should not be seen as a ground truth, but rather as an independent temperature data set to compare with. 
While the 20CRv3.SI reanalysis has been shown to perform reasonably even in the 19th century, there are known biases in the 20CRv3.SI product that should be treated with caution \cite{slivinski_evaluation_2021}.

We also evaluate our reconstruction method on the more recent HadCRUT5 \cite{morice_updated_2021} dataset, which has further improved coverage in recent years compared to earlier versions. 
We choose a late month of the HadCRUT5 data that has no gaps, here January 2021, apply all HadCRUT4 derived masks and subsequently reconstruct the artificially masked temperature field (fig.~\ref{fig:hc5_2052_map}).
Generally, LaMa is able to reconstruct the temperature fields well compared to the ground truth. 
Even for the mask with the largest RMSE, we find good agreement of the spatial temperature patterns with the ground truth (fig.~\ref{fig:hc5_2052_map}B).
LaMa is able to reconstruct the spatial patterns in most parts of the Pacific, America, and Eurasia.
However, there are notable difference between ground truth and reconstructions in the polar regions, with considerably underestimated Arctic temperatures in the reconstructions. 
For the mask that leads to the lowest RMSE, most masked grid cells are reconstructed very well, including the polar regions (fig.~\ref{fig:hc5_2052_map}C). 
On average, the site-wise RMSE shows the largest differences in the polar regions, northern Asia, and northern North America (Fig.~\ref{fig:hc5_2052_map}D).
We calculate the difference between the ground truth GMT and the reconstructed GMT for each mask  (fig.~\ref{fig:hc5_2052_ts}). 
The reconstructed GMT shows the largest difference for masks with a coverage of more than 70\%, with a maximum difference of -0.27$^\circ$C. 
For the early masks, the reconstructed GMT is generally lower than the ground truth GMT.
For later masks, the reconstructed GMT is close to the ground truth GMT.
The average difference between the ground truth GMT and the reconstructed GMT over all masks is close to 0$^\circ$C (fig.~\ref{fig:hc5_2052_ts}).

\subsection*{Beyond HadCRUT}
LaMa is able to generalize to higher resolutions than the ones it is trained on and is not restricted to temperature fields.
We inpaint the 90x90~pixel Berkeley Earth Surface Temperatures (BEST) \cite{rohde_berkeley_2020} using the LaMa model trained on the 72x72~pixel CMIP5 images (fig.~\ref{fig:best_maps}). 
We also show an example reconstruction of sea ice concentration to show the application to a structurally different climatic field (fig.~\ref{fig:sea_ice}).

We do not modify the trained model before evaluating on the BEST temperature records.  
We transform the BEST temperature records to 90x90~pixel images with the same procedure as before, which corresponds to a 156.25\% higher resolution than the images we trained the model on.
LaMa, trained solely on the HadCRUT4 masks, shows visible artifacts in the inpainted spatial fields, especially at the edges of the gaps (fig.~\ref{fig:best_maps_artifcats}), and is therefore not suitable for reconstructions on unseen masks. 
This problem can be easily facilitated by employing a different mask generation algorithm during training.
By generating random masks during training on the fly, using a previously proposed mask generation algorithm \cite{suvorov_resolution-robust_2021}, LaMa generalizes to different masks than the ones seen during training.
In the following, we call LaMa with randomly generated masks during the training \textit{LaMa random}. 
We show that LaMa random is able to inpaint the missing areas in the BEST record without any strong artifacts (fig.~\ref{fig:best_maps}). 
LaMa random is therefore better suited for generalization tasks when the final mask shapes for inference are not known during the training. 

It should be noted that the very low resolution we use for the training images limits the application to higher resolutions. 
However, it has been shown that LaMa can generalize to resolutions up to four times higher than those it is trained on \cite{suvorov_resolution-robust_2021}. Hence, a significantly stronger upsampling than from CMIP5 (72x72 pixels) to BEST (90x90 pixels) should be possible. 

To show the applicability to a different climate field, we train LaMa on the daily sea ice concentration from 1979-2022, taken from the ERA5 reanalysis \cite{bell_era5_2021, era5_sea_ice} with a resolution of 180x1440~px (Northern Hemisphere).
This gives a total of 15,450 monthly fields, where we hold out 1,054 random months for evaluation and 1,043 random months for validation.
Even for this relatively small training sample, LaMa is able to reconstruct the spatial extent and concentration of the sea ice reasonably well (fig.~\ref{fig:sea_ice}).
LaMa learns the continent distribution during training and correctly predicts the extent of the sea ice, given very little information from the unmasked areas (fig.~\ref{fig:sea_ice}A,D). 
LaMa correctly reconstructs the seasonality of the sea ice concentration with a maximum in winter (fig.~\ref{fig:sea_ice}a) and a minimum in summer (fig.~\ref{fig:sea_ice}D).
The largest deviations from the ground truth are generally at the edges of the sea ice, while the central Arctic shows the lowest error (fig.~\ref{fig:sea_ice}C).  
This example case of the sea ice concentration shows that LaMa can be applied to a variety of structurally different climate fields. 

\section*{Discussion}
Reconstruction of historical observations is an active and important research field in climate science with vast implications for the present climate, short- and long-term future projections, and climate change attribution. 
Previously used methods often struggle with large irregular gaps in climate fields or with resolving spatial patterns. 
We show that LaMa is able to realistically reconstruct global temperature records across different data sets and resolutions.
LaMa clearly outperforms the widely used kriging with a 21.0\% smaller spatially averaged site-wise RMSE (Tab.~\ref{tab:comparison}). 
Furthermore, our method outperforms a previously proposed deep learning method based on partial convolutions \cite{kadow_artificial_2020}. 
In terms of the spatially averaged site-wise RMSE, LaMa outperforms PConv by 13.5\% (Tab.~\ref{tab:comparison}), and 78\% of the grid cells show a lower site-wise RMSE on the test set (Fig.~\ref{fig:Fig_3}F).
We note that approaches using Principal Component Analysis (PCA) have been successfully used to reconstruct temperature records \cite{beckers_eof_2003, wang_reconstructed_2020}.
In particular, PCA has been shown to be able to reconstruct the El Niño-Southern Oscillation remarkably well \cite{smith_reconstruction_1996}. 
While generally PCA shows a better performance than kriging in reconstructing spatial temperature patterns, PCA shows overall worse performance than PConv \cite{kadow_artificial_2020}.
More specifically, a thorough comparison of the PConv method with a PCA-based reconstruction shows that PCA introduces additional biases when applying cross-validation, and is generally outperformed by PConv \cite{kadow_artificial_2020}. 
Since, as we have shown, our approach generally outperforms the PConv method, it is thus clear that it also gives better reconstructions than PCA.

A substantial advantage of LaMa over PConv and other, statistical methods such as kriging, is the global context provided by Fast Fourier Convolutions (FFC), in contrast to the local context of PConv. 
For climate fields with global teleconnections, such as temperature fields, the FFC blocks in LaMa allow for a global context, promising more accurate spatial reconstructions.
We train our model on rather low-resolution images (72x72~pixel), which makes it difficult to resolve global teleconnections.   
Nonetheless, our model is able to realistically reconstruct spatial temperature patterns on a global level. 
Training on high-resolution images is limited by the available GPU infrastructure. 
However, due to the ability of LaMa to train on lower-resolution images than the ones it is evaluated on, this problem can be mitigated. 
While the training of the model can take several wall time days, depending on the training size, the evaluation is done in the order of minutes. 
It is hence much faster than gap-filling with dynamical models and still faster than statistical methods.

While LaMa can generalize to data different from its training set, the masks derived from BEST appear to be too dissimilar from the HadCRUT4 masks, which LaMa was trained on, to yield sensible reconstructions.
We find that randomly generated masks during training ensure applicability on masks never seen during training, but on the HadCRUT4-derived masks LaMa almost consistently outperforms LaMa random in terms of the error metrics (fig.~\ref{fig:site_wise_difference_lama}).
We note, however, that an improved random mask generation during training could potentially further improve LaMa random.
Furthermore, we show that LaMa can be applied to a variety of structurally very different climate fields (fig.~\ref{fig:sea_ice}).

The global time series reconstructed by LaMa shows slightly lower GMT until the year 1880~AD.
With LaMa we find a GMT of $1.09^\circ$C above pre-industrial level (1850-1900) for the period 2010-2020, while the best estimate based on the masked HadCRUT4 data set gives a warming level of $1.00^\circ$C for the same period.
This is mostly due to a colder Southern Hemisphere reconstructed by LaMa, especially in the Antarctic region.
Due to the sparse observations at the poles, it is difficult to validate the plausibility of the reconstructed temperatures in these areas. 
While the colder Antarctica reconstructed by LaMa is a priori not implausible, LaMa shows worse performance in larger parts of Antarctica than PConv and kriging (Fig.~\ref{fig:Fig_3}D,F). 
Additionally, the evaluation on the HadCRUT5 month shows that LaMa underestimates the GMT in the first 300 to 400 masks, suggesting that our approach generally underestimates the GMT in the beginning of the reconstruction.  
We attribute a possible underestimation of Antarctic temperatures to two reasons.
Firstly, the variability of the surface temperature in Antarctica across the CMIP5 ensemble is large \cite{tang_assessment_2018}. 
For a similar global surface temperature distribution, the temperatures in Antarctica might differ vastly between the single ensemble members. 
This makes it hard for the ML model to learn useful spatial connections that lead to reasonable Antarctic temperature predictions. 
Secondly, the inpainting problem turns into an outpainting problem for the polar regions, which is inherently harder.
There are almost no known measurements for any time step in close proximity to Antarctica. 
While LaMa is able to extrapolate the polar regions reasonably well, the performance is worse than in the non-polar regions (Fig.~\ref{fig:Fig_3}).
However, the left and right edges of the images, corresponding to the prime meridian, do not necessarily show a higher RMSE than other regions (Fig.~\ref{fig:Fig_3}), which suggests that this is not the main reason.  

For a masking ratio of more than 80\%, LaMa shows a higher mean site-wise RMSE on the testing set than the other methods (Fig.~\ref{fig:Fig_4}D). 
This is due to artifacts with unusually high temperatures in western Antarctica (bottom right corner of the image) for some of the CMIP5 ensemble members. 
Similarly to Kadow et al. \cite{kadow_artificial_2020}, we attribute this to effects at the edges of the images, as mentioned above. 
LaMa random does not exhibit these artifacts, which suggests that a more sophisticated mask generation during training could resolve these issues. 
We do not observe any artifacts in the other reconstructed data sets.

Convolutional neural networks (CNN) are generally not able to capture the spherical geometry of the Earth well, which can lead to the aforementioned artifacts at the edges and worse performance in the polar regions.
Graph neural networks or spherical convolutions could facilitate this problem \cite{kadow_artificial_2020, keisler_forecasting_2022, scher_physics-inspired_2023}.
First attempts of applying CNNs with spherical harmonics as basis functions to weather forecasting show promising results \cite{esteves_scaling_2023}. 
However, spherical CNNs remain to be computationally much more expensive than regular convolutional networks \cite{esteves_scaling_2023}.
Alternatively, one could train several different models for different, smaller regions of the Earth surface using corresponding geographical projections, and merge the resulting output.
This would ameliorate the problem of the model not being aware of the changing grid cell size and hence correlation towards the poles in a regular projection.
However, besides the increased computational demand, this is only feasible if the training data is available in appropriate projections, or if the raw observational/model data is available so that the different projections can be carried out. 
Otherwise, interpolation artifacts from re-projecting would lead to additional errors.
The rising popularity of generative models makes it a promising alternative to CNN-based models for the reconstruction of climate fields.
Especially, recently introduced diffusion models \cite{lugmayr_repaint_2022,wei_diffusion_2023} show promising performance on image inpainting tasks. 
Furthermore, by using video-inpainting techniques \cite{hoppe_diffusion_2022} rather than image-inpainting, the temporal dimension of the data could directly be taken into account.
However, little work has been done so far in that direction and the physical plausibility of such models remains uncertain. 

Reconstructions via deep learning can aid in understanding past and present changes in the Earth system. 
By learning the spatiotemporal patterns of the underlying climate fields, LaMa is able to realistically reconstruct a variety of observables with varying resolutions. 
Our easy-to-use deep learning model clearly outperforms previous methods and therefore serves as an alternative to conventional methods used in the geosciences. 

\section*{Methods}
\subsection*{Resolution-robust Large Mask Inpainting with Fourier Convolutions (LaMa)}
We utilize the recently introduced LaMa model \cite{suvorov_resolution-robust_2021} that builds on Fast Fourier Convolutions (FCC) for reconstructing missing image regions. 
LaMa is a feed-forward ResNet-like inpainting network with a multi-component loss. 
LaMa has been shown to outperform other ML methods such as Aggregated Contextual Transformations for High-Resolution Image Inpainting (AOT GAN) \cite{zeng_aggregated_2021}, Image Inpainting with Learnable Feature Imputation \cite{hukkelas_image_2020}, or latent diffusion models \cite{suvorov_resolution-robust_2021, rombach_high-resolution_2022, kulshreshtha_feature_2022}, and is able to inpaint large missing areas with a high receptive field. 

LaMa is designed to inpaint masked images. 
Given an image $x$ and a binary mask $m$, the input is prepared by stacking the masked image with the mask itself: $x' = \texttt{stack}(x \odot m, m)$, where $\odot$ denotes element-wise multiplication. This results in a 4-channel input tensor.
The core of LaMa is the FFC block, which utilizes channel-wise Fast Fourier Transforms (FFT). 
FFC's primary advantage is its ability to capture global context in early network layers, unlike conventional convolutions with limited spatial receptive fields \cite{NEURIPS2020_2fd5d41e}.
An FFC block consists of two interconnected branches:
\begin{itemize}
    \item A local branch in the spatial domain using conventional convolutions
    \item A global branch in the spectral domain using real FFT
\end{itemize}
The input tensor $x$ is split along feature channels into local ($x^l$) and global ($x^g$) components:
\begin{itemize}
    \item $x^l \in \mathbb{R}^{H \times W \times (1-\alpha)C}$ captures small-scale information
    \item $x^g \in \mathbb{R}^{H \times W \times \alpha C}$ captures large-scale context
\end{itemize}

Here, $H \times W$ represents spatial dimensions, $C$ is the number of channels, and $\alpha \in [0,1]$ is the hyperparameter determining the channel split ratio between global and local branches.
The FFC block's output $y = \{y^l, y^g\}$ maintains the same dimensions and split ratio as the input. The block's operations can be described as:
\begin{align}
y^l &= y^{l \to l} + y^{g \to l} = f_l(x^l) + f_{g \to l}(x^g) \\
y^g &= y^{g \to g} + y^{l \to g} = f_g(x^g) + f_{l \to g}(x^l)
\end{align}
Where $f_l$ is the local operation via conventional convolutions, $f_g$ is the global spectral transformer and $f_{g \to l}$ and $f_{l \to g}$ denote the inter-path transitions via conventional convolutions.
The spectral transform $f_g$ in the global path ensures global context in early layers. 
While the global branch can suffice for realistic inpainting, the local branch adds network stability \cite{yannic_kilcher_resolution-robust_2021}.
FFC's image-wide receptive field allows for global context in early network layers. 
Conventional convolutions would require substantially more layers to achieve a similar receptive field, increasing computational demands.

The multi-component loss of LaMa is the second crucial part for achieving realistic inpainting results.
The full loss function $\mathcal{L}$ is given by
\begin{align}
    \mathcal{L} = \kappa L_\text{adv}+\alpha \mathcal L_{\text{HRFPL}} + \beta \mathcal{L}_\text{DiscPL} + \gamma R_1
\end{align}
with an adversarial loss $L_\text{adv}$, the high receptive field perceptual loss $L_{\text{HRFPL}}$, a discriminator-based perceptual loss or feature-matching loss $\mathcal{L}_\text{DiscPL}$ and a regularization term $R_1$.
The adversarial loss $L_\text{adv}$ ensures that the inpainting network $f_\Theta(x')$ generates natural-looking local details. 
The high receptive field perceptual loss $L_{\text{HRFPL}}$ evaluates the similarity between the target and predicted image using a pre-trained network, here a pre-trained ResNet50 network with dilated convolutions. 
The discriminator-based perceptual loss increases stability and performance \cite{suvorov_resolution-robust_2021}. 
The regularization term prevents overfitting and allows better generalization.
The weights are given by $\kappa=10$, $\alpha=30$, $\beta=100$ and $\gamma=0.001$. 

LaMa is able to inpaint high-resolution images even if trained on lower-resolution images. 
We extend and modify the model to allow drawing of pre-generated masks from Hierarchical Data Format version 5 (HDF5) files during training and evaluation, as well as to enable training and evaluation on rectangular images.
LaMa outperforms the method based on partial convolutions \cite{kadow_artificial_2020} in terms of spatial metrics and shows comparable performance on a temporal mean global scale. 
For the full description of the network architecture we refer to the original paper \cite{suvorov_resolution-robust_2021}.

\subsection*{Training procedure, preprocessing \& validation}
\subsubsection*{Training strategy}
We train the model on the monthly surface temperature (tas) of 239,616 CMIP5 ensemble members following \cite{kadow_artificial_2020}. 
For the training and evaluation, we transform the temperature records into 72x72~px greyscale png-images with three identical RGB channels.   
We normalize the images with respect to the maximum and minimum values in the full CMIP5 set such that the maximum temperature corresponds to 255 and the minimum value to 0. 
Therefore, the maximum resolution of the reconstruction is given by $\frac{\left|\mathbf{u}_\text{max}\right|+\left|\mathbf{u}_\text{min}\right|}{255}$ with $\mathbf{u}$ as the climatic field of interest. 
For our monthly mean temperature reconstruction, this leads to an effective maximum resolution of approximately $0.19\,^\circ$C.
We convert the floats to integers during the transformation of the temperature records to images by truncation. 
In the following, we take the transformed temperature records as ground truth when we refer to the metrics of our ML model. 
For simplicity, we only plot the non-transformed HadCRUT4 temperature time series and add the difference between the HadCRUT4 masked temperature records and the transformed HadCRUT4 temperature records when we plot the reconstructed LaMa temperature time series.  
We hold out 27,702 images for validation and 2,250 images for evaluation.
Furthermore, we hold out one CMIP5 ensemble member from training for comparison with the approach by \cite{kadow_artificial_2020}.
We use two different mask-generation methods during the training. 
For the first approach, we train the model on randomly drawn masks derived from the HadCRUT4 \cite{morice_quantifying_2012} missing masks. 
In the following, we name this model \textit{LaMa}. 
Our second approach generates random masks following the approach of \cite{suvorov_resolution-robust_2021} during training, hereafter \textit{LaMa random}. 

To show the independence of our test set from the training set, we retrain our model on an alternative training-validation-test split and repeat the error analysis.
We withhold 17 random years from the training set, instead of withholding every 9th month from the training set. 
By withholding a year from the training set, we mean that the whole year for all 144 CMIP5 ensemble members is removed from the training set.  
We use 15 of these 17 years for validation (corresponding to 25670 monthly fields) and 2 years for testing (corresponding to 3456 monthly fields).
We train the model for 55 epochs and otherwise follow the same training strategy as mentioned above.

\subsubsection*{Random mask generation}
The random mask generation algorithm works as follows.
With a 77\% probability, several line segments are drawn. 
A random starting point on the image is chosen where multiple line segments (we choose minimum 4 and maximum 5) are drawn, which correspond to the masked area. 
Each of the line segments has a random angle, length (max. 40~px), and width (max. 35~px).
With 23\% probability, random rectangle boxes with a minimum size of 10\,px and a maximum size of 70\,px (minimum 1 box and maximum 5 boxes) are drawn instead of line segments.
Examples of some random masks generated during the training are shown in fig.~\ref{fig:random_masks}.

\subsubsection*{Hardware}
We train both models on two NVidia Tesla V100 GPUs or alternatively on one NVIDIA H100 Tensor Core GPU with a maximum of 60 epochs with a batch size of 100. 
We choose the training checkpoint with the lowest error metrics on the evaluation data set for each model. 
Therefore, we use the 57th checkpoint for LaMa random and the 60th checkpoint for LaMa fixed. 
For the full training parameters, we refer to the configuration files in the GitHub repository.

\subsubsection*{Evaluation}
As main evaluation metrics, we use the spatial RMSE and the site-wise RMSE.
We refer to the square root of the spatially weighted average of the squared differences between the ground truth and the inpainted image as the spatial RMSE:
\begin{align}
    \text{RMSE} = \sqrt{\frac{\sum_{i=1}^{n} w_i (y_i - \hat{y}_i)^2}{\sum_{i=1}^{n} w_i}}
\end{align}
with the inpainted grid cell $y_i$ and the ground truth $\hat{y}_i$ at the given time steps, the number of grid cells $n$ and the weights $w_i= \cos(\phi_i)$ that are given by the latitude $\phi_i$ of the corresponding point $y_i$, which corresponds to the area of the grid cell for a regular rectangular grid.
Additionally, we define the site-wise RMSE as the RMSE at each site (i.e., grid cell), averaged over the time dimension. 
The mean site-wise RMSE is the spatially weighted average of the site-wise RMSE in all grid cells. 
Hence, the main difference between the two main evaluation metrics that we focus on is the order of averaging in time and space.
In each time step, we exclude the grid cells that have known values for the calculations of the RMSE.  
It is explicitly stated if we show a non-weighted RMSE.

In addition to the described validation procedures, we compare our reconstructions visually with the 20CRv3.SI reanalysis \cite{compo_twentieth_2011}. 
20CRv3.SI is a reanalysis, or four-dimensional weather reconstitution, spanning the years 1836 to 2015.
The reanalysis product only assimilates surface observations of synoptic pressure into the National Oceanic and Atmospheric Administration's (NOAA) Global Forecast System and prescribe sea surface temperature and sea ice distribution to estimate other climate variables. 

\newpage


\clearpage 

%
\bibliography{references} 
\bibliographystyle{sciencemag}

%
%
%
%
%
%


\paragraph{Data and Materials Availability} All data needed to evaluate the conclusions in the paper are present in the paper and the Supplementary Materials. The HadCRUT4 data \cite{morice_quantifying_2012} is available at \url{https://www.metoffice.gov.uk/hadobs/hadcrut4/}. The HadCRUT5 data \cite{morice_updated_2021} is available at \url{https://www.metoffice.gov.uk/hadobs/hadcrut5/data/HadCRUT.5.0.2.0/download.html}. 
The NOAA/CIRES/DOE 20th Century Reanalysis (V3) data is provided by the NOAA PSL, Boulder, Colorado, USA, and is available at their website at \url{https://psl.noaa.gov/data/gridded/data.20thC_ReanV3.html} \cite{slivinski_towards_2019}. 
The BEST data set is freely available at \url{https://berkeleyearth.org/data/} \cite{rohde_berkeley_2020}. 
The sea ice concentration data is available at \url{https://cds.climate.copernicus.eu/cdsapp##!/dataset/satellite-sea-ice-concentration}. 
The model check points as well as the reconstructed fields shown are available on zenodo at \url{https://www.doi.org/10.5281/zenodo.10512175}.
The original LaMa model is available on GitHub at \url{https://github.com/advimman/lama}. Our modified version is available at \url{https://github.com/NilsBochow/lama_reconstruction}. 
Both LaMa versions are also archived on zenodo at \url{https://www.doi.org/10.5281/zenodo.10512175}. We use the software package PyKrige \cite{murphy_geostat-frameworkpykrige_2022} for reconstructing the temperature records via kriging. 
\paragraph{Author contributions} N.Bow. and N.Boe. conceived and designed the study. N.Bow. carried out the computations and analyzed the results with help from A.P. N.Boe. and M.R. supervised the project. All authors discussed the results. N.Bow. wrote the paper with contributions from all authors.
\paragraph{Competing interests} The authors declare no competing interests.
\paragraph{Funding} N.Boe. acknowledges funding from ClimTip which has received funding from the European Union's Horizon Europe research and innovation program under Grant Agreement No. 101137601. N.Boe. acknowledges further funding by the Volkswagen foundation and the European Union’s Horizon 2020 research and innovation program under the Marie Sklodowska-Curie grant agreement No. 956170, as well as from the Federal Ministry of Education and Research under grant No. 01LS2001A. M.R., A.P. and N.Bow. were supported by the UiT Aurora Centre Program, UiT The Arctic University of Norway (2020), and the Research Council of Norway (project number 314570). 
\paragraph{Acknowledgments} This is ClimTip contribution \#19; the ClimTip project has received funding from the European Union's Horizon Europe research and innovation programme under grant agreement No. 101137601. 
Parts of the computations were performed on resources provided by Sigma2 - the National Infrastructure for High Performance Computing and Data Storage in Norway under the project nn8008k.
Support for the Twentieth Century Reanalysis Project version 3 data set is provided by the U.S. Department of Energy, Office of Science Biological and Environmental Research (BER), by the National Oceanic and Atmospheric Administration Climate Program Office, and by the NOAA Earth System Research Laboratory Physical Sciences Laboratory. The authors gratefully acknowledge the European Regional Development Fund (ERDF), the German Federal Ministry of Education and Research and the Land Brandenburg for supporting this project by providing resources on the high performance computer system at the Potsdam Institute for Climate Impact Research. 
Parts of the publication charges for this article have been funded by a grant from the publication fund of UiT The Arctic University of Norway. Some of the figures are made with color maps by Crameri et al. \cite{crameri_misuse_2020}. We thank R. Suvorov for answering technical questions regarding their LaMa model, J. Meuer for providing the CMIP5 training and validation data and C. Kadow for providing the output of their model.


\subsection*{Supplementary materials}

Figs. S1 to S12\\
\newpage

 \clearpage

\clearpage

\clearpage

\clearpage

\begin{table}[t]

\begin{tabular}{l*{1} {|p{4cm}p{4cm}}}
    \midrule
    Model  &  spatially averaged site-wise RMSE [$^\circ$C] & temporally averaged spatial RMSE [$^\circ$C]  \\
    \bottomrule
    LaMa        & 0.64 (21.0\%) & 0.99 (16.8\%)  \\
    
    LaMa random     & 0.74 (8.6\%) & 1.10 (7.6\%) \\
    
    PConv    & 0.74 (8.6\%)& 1.08 (9.2\%)  \\
    
    Kriging      & 0.81 & 1.19  \\
    
    \bottomrule
\end{tabular} 
\caption{\textbf{Comparison between all methods in terms of the RMSE.} Spatially weighted average of the site-wise RMSE and average spatial RMSE for both LaMa methods, PConv, and kriging for the single CMIP5 ensemble member. The improvement compared to kriging is denoted in the parentheses. LaMa shows considerable improvement in comparison with the other methods. LaMa random outperforms kriging and has a similar performance as PConv. }
\label{tab:comparison}
\end{table}


\clearpage


\renewcommand{\thefigure}{S\arabic{figure}}
\renewcommand{\thetable}{S\arabic{table}}
\renewcommand{\theequation}{S\arabic{equation}}
\renewcommand{\thepage}{S\arabic{page}}
\setcounter{figure}{0}
\setcounter{table}{0}
\setcounter{equation}{0}
\setcounter{page}{1} 


\begin{center}
\section*{Supplementary Materials for\\ \scititle}

	Nils~Bochow$^{\ast}$,
	Anna~Poltronieri,
	Martin~Rypdal,
    Niklas~Boers \\
\small$^\ast$Corresponding author. Email: contact@nilsbochow.com

\end{center}

\subsubsection*{This PDF file includes:}

Figures S1 to S12\\

\newpage


\begin{figure}[t]
\includegraphics[width=\textwidth]{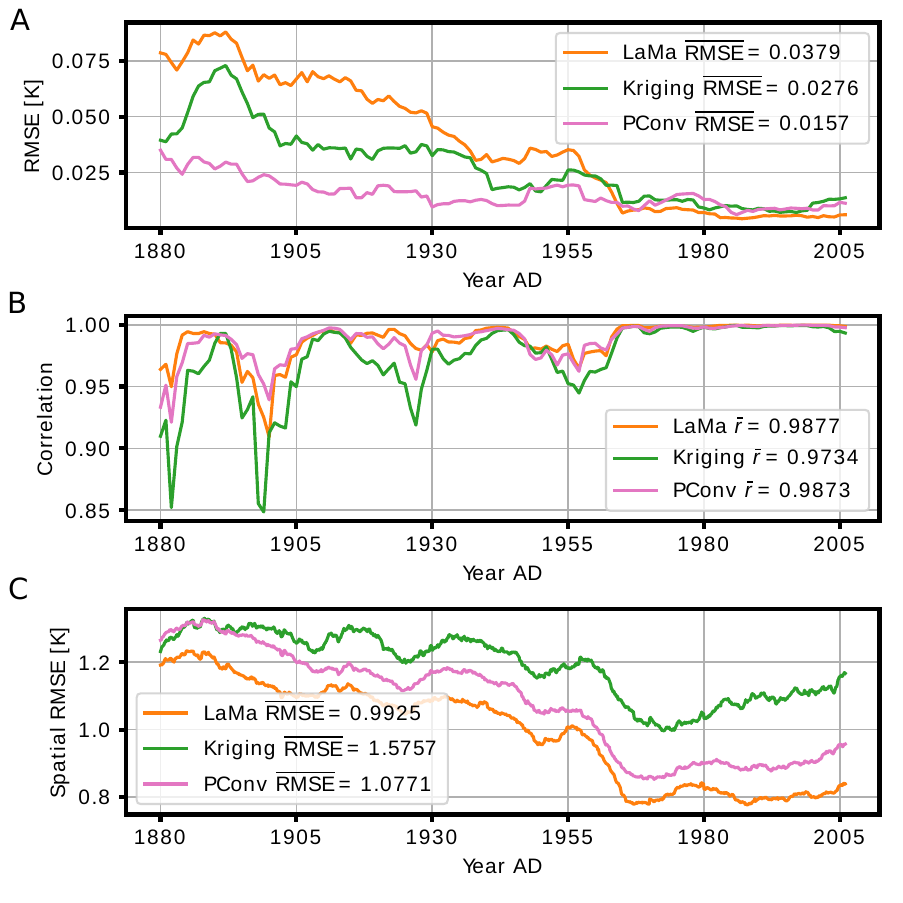}
\caption{\textbf{Comparison with reconstruction method via PConv and kriging for held-out CMIP5 member.} \textbf{(A)} Root-mean-squared error between infilled yearly time series for LaMa, kriging and PConv \cite{kadow_artificial_2020} and the held out CMIP5 member (ground truth) in a rolling window with size $w=10$\,years (1870-2005~AD). The mean RMSEs over the whole time period are denoted in the legend. \textbf{(B)} Same as \textbf{(A)} but for correlation between yearly temperature time series. \textbf{(C)} Weighted spatial root-mean-squared error of monthly temperature fields in a rolling window with window size $w=10$\,years.}
\label{fig:kadow_timeseries_comparison}
\end{figure}

\begin{figure}[t]
\includegraphics[width=0.7\textwidth]{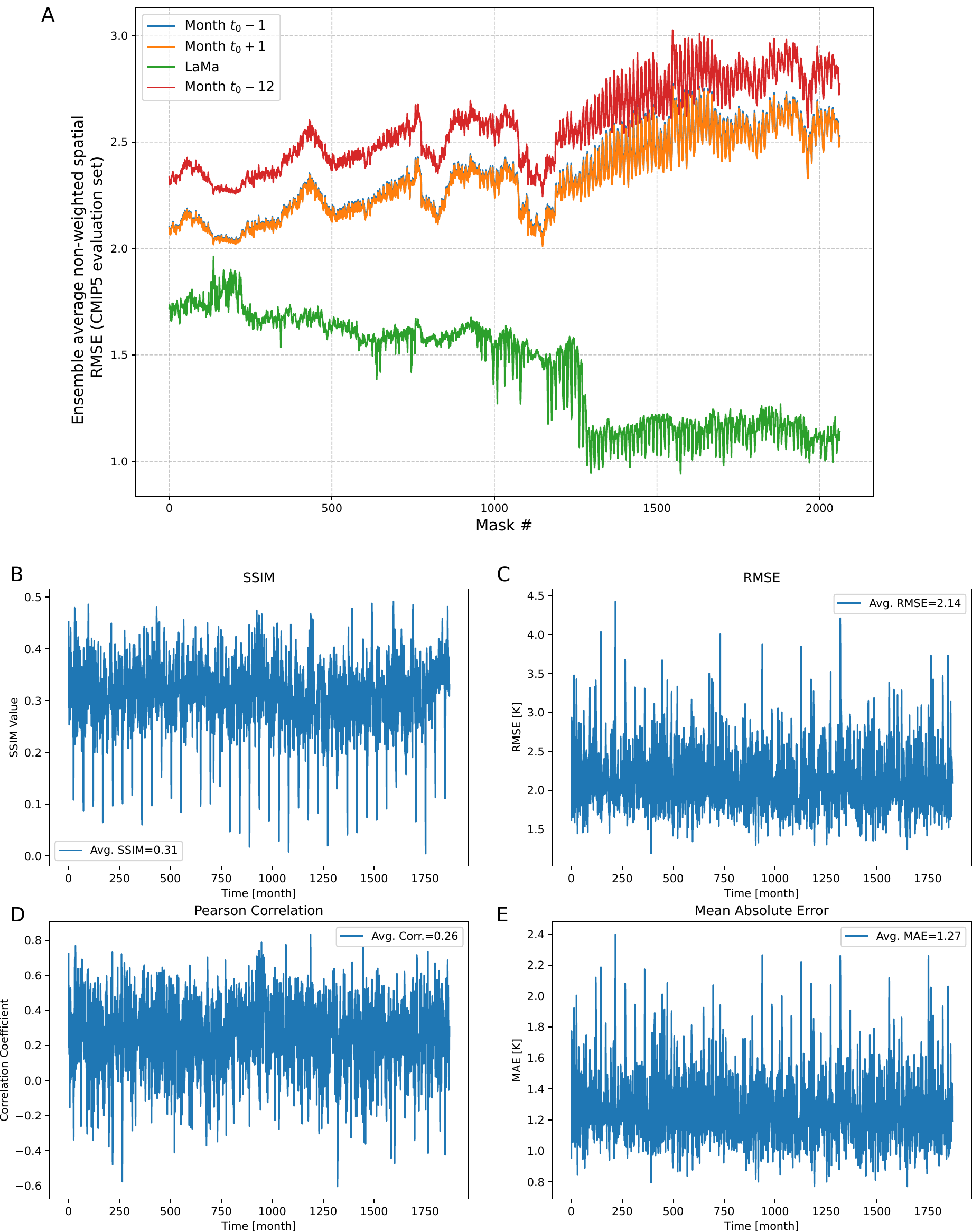}
\caption{\textbf{Analysis of similarity of CMIP5 evaluation data set.} \textbf{(A)} Ensemble mean of the spatial root-mean-squared error (non-weighted) of the reconstruction via LaMa and the naive approach of filling in the missing grid cells with the previous month (Month $t_0-1$), the following month (Month $t_0+1$) and the same month of the previous year (Month $t_0-12$) on the CMIP5 evaluation set. We calculate the spatial RMSE of each ensemble member (2251) average for each mask. The RMSE of the LaMa reconstruction is generally lower than of the naive approaches. \textbf{(B)} Structural similarity index measure (SSIM) between consecutive months of one model of the CMIP5 ensemble. A SSIM of 1 indicates perfect similarity between two images, while 0 corresponds to no similarity \cite{wang_mean_2009}. The average metric is depicted in the legend. \textbf{(C,D,E)} Same as \textbf{B} but for the RMSE, Pearson correlation and MSE, respectively. Generally, the similarity between consecutive months is low.}
\label{fig:similarity_metrics}
\end{figure}

\begin{figure}[t]
\includegraphics[width=0.9\textwidth]{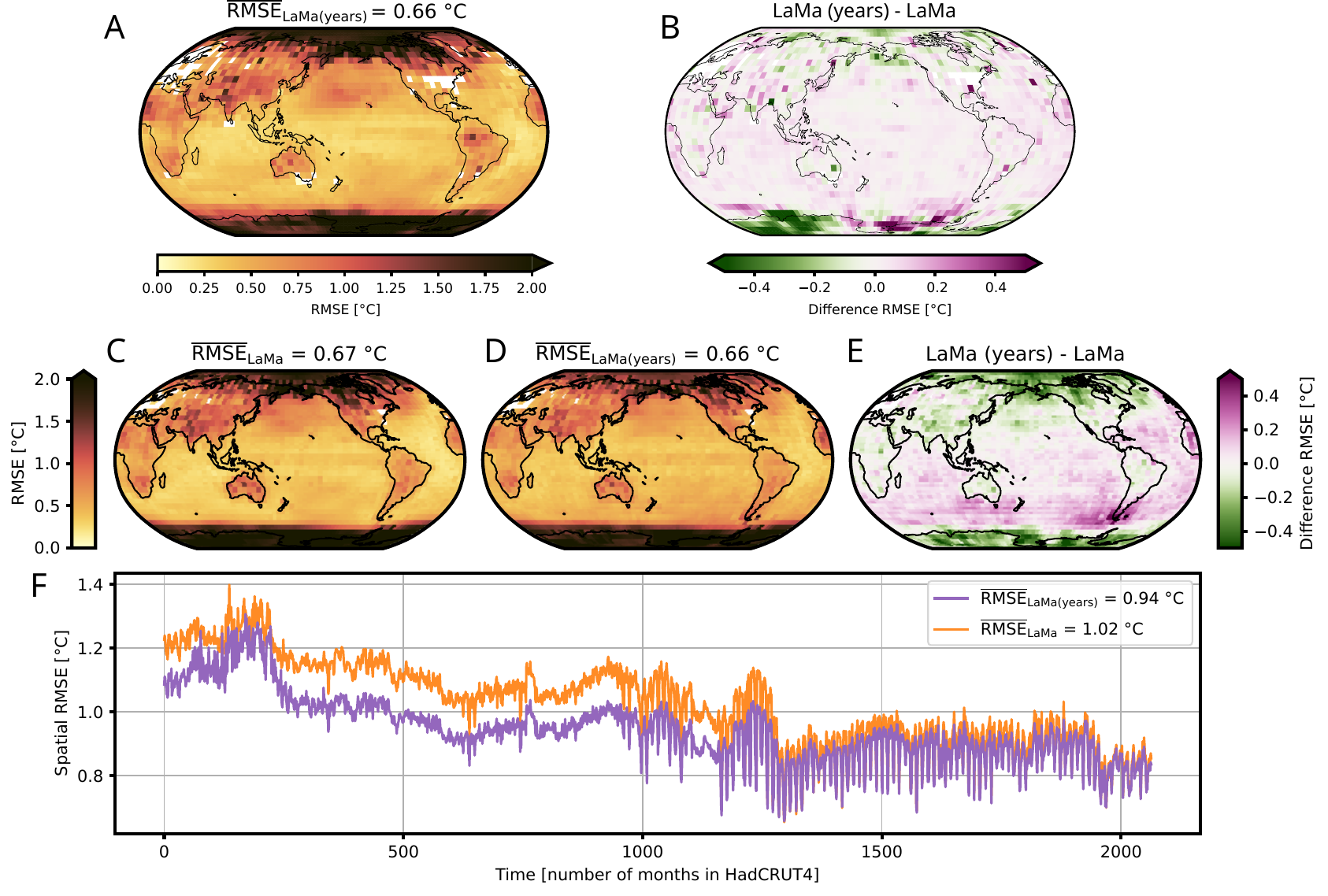}
\caption{\textbf{Error metrics for LaMa trained on alternative train-validation-test split.} \textbf{(A)} Site-wise RSME for LaMa trained by withholding random years instead of every 9th month (named LaMa (years)) on the 145th CMIP5 ensemble member. \textbf{(B)} Difference in site-wise RMSE between LaMa (years) and LaMa shown in the main text on the 145th CMIP5 ensemble member from 1870-2005. Green areas denote regions where LaMa (years) has a lower RMSE than LaMa. \textbf{(C)} Site-wise RMSE of LaMa on the 2251 withheld months (test set) as shown in the main text. \textbf{(D)} Site-wise RMSE of LaMa (years) on the alternative test set (3456 monthly fields). \textbf{(E)} Difference in site-wise RMSE between the two models. Green areas denote regions where LaMa (years) has a lower RMSE than LaMa. Note that the two test sets are not the same. \textbf{(F)} Spatial RMSE for both methods and all HadCRUT4 masks, which are ordered in time; note that generally the size of the masks in terms of the number of missing data declines over time. LaMa (years) shows a lower spatial RMSE for almost all masks.}
\label{fig:alternative_test_set}
\end{figure}

\begin{figure}[t]
\includegraphics[width=\textwidth]{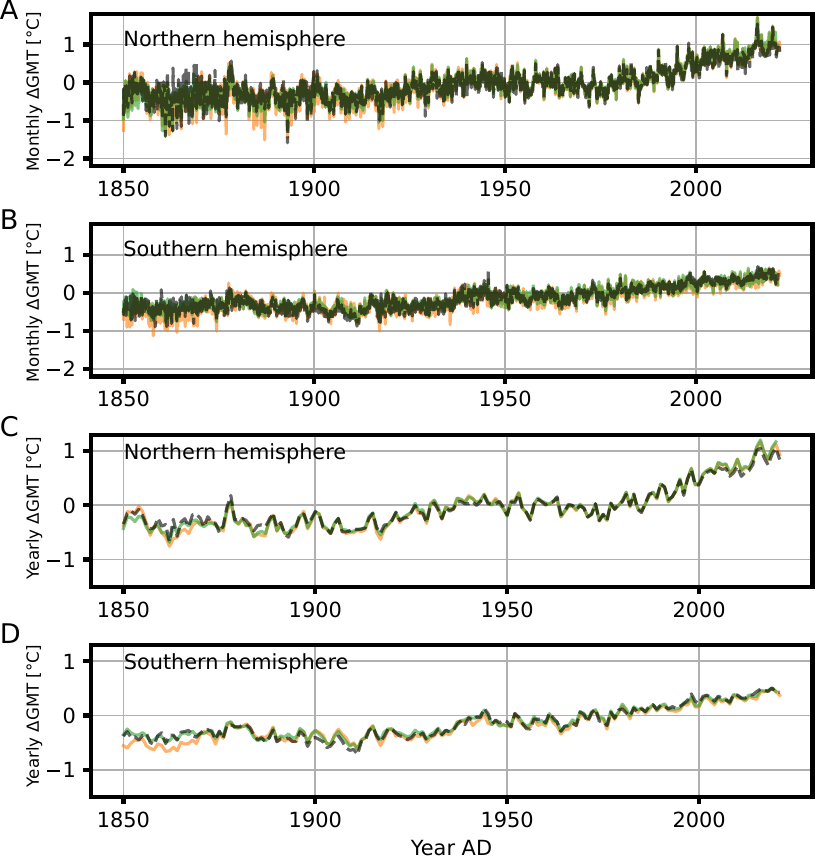}
\caption{\textbf{Reconstructed HadCRUT4 global mean temperature time series for Northern and Southern Hemisphere.} \textbf{(A)} Monthly HadCRUT4 timeseries from 1850 to 2022 for the Northern Hemisphere relative to the period 1960-1990. The reconstructions from LaMa and kriging as well as the mean of the masked HadCRUT4 records are shown. The dashed black curve is the spatially averaged GMT derived from the incomplete HadCRUT4 observations. \textbf{(B)} Same as \textbf{A} but for the Southern Hemisphere. \textbf{(C)} Same as \textbf{A} but for the yearly averaged GMT. Additionally, we show the reconstruction based on the PConv method. \textbf{(D)} Same as \textbf{C} but for the Southern Hemisphere.}
\label{fig:ts_hemispheres}
\end{figure}

\begin{figure}[t]
\includegraphics[width=\textwidth]{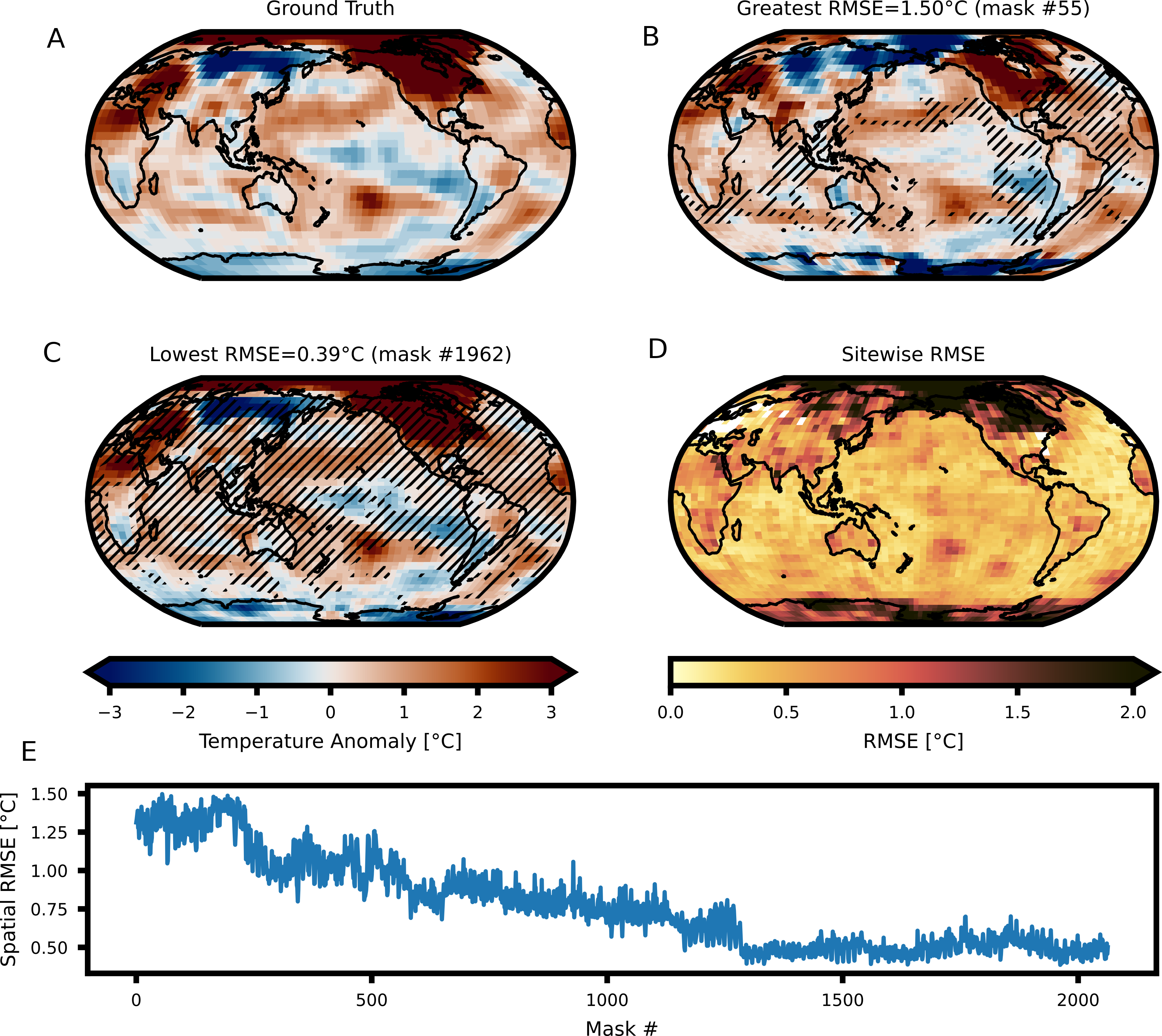}
\caption{\textbf{Reconstructed HadCRUT5 temperature field for January 2021 with HadCRUT4 derived masks.} \textbf{(A)} Original HadCRUT5 temperature anomaly for January 2021 (ground truth). \textbf{(B)} Reconstructed temperature anomalies with LaMa for the month with the highest spatially weighted spatial root-mean squared error. The hatched area denotes the grid cells that are unmasked, i.e. the information fed into LaMa. \textbf{(C)} Same as \textbf{B} but for the month with the lowest RMSE. \textbf{(D)} Mean site-wise RMSE for the all applied masks. White areas denote grid cells with information available throughout all masks. \textbf{(E)} Time series of the spatially weighted spatial RMSE for all applied masks. }
\label{fig:hc5_2052_map}
\end{figure}
\clearpage

\begin{figure}[t]
\includegraphics[width=\textwidth]{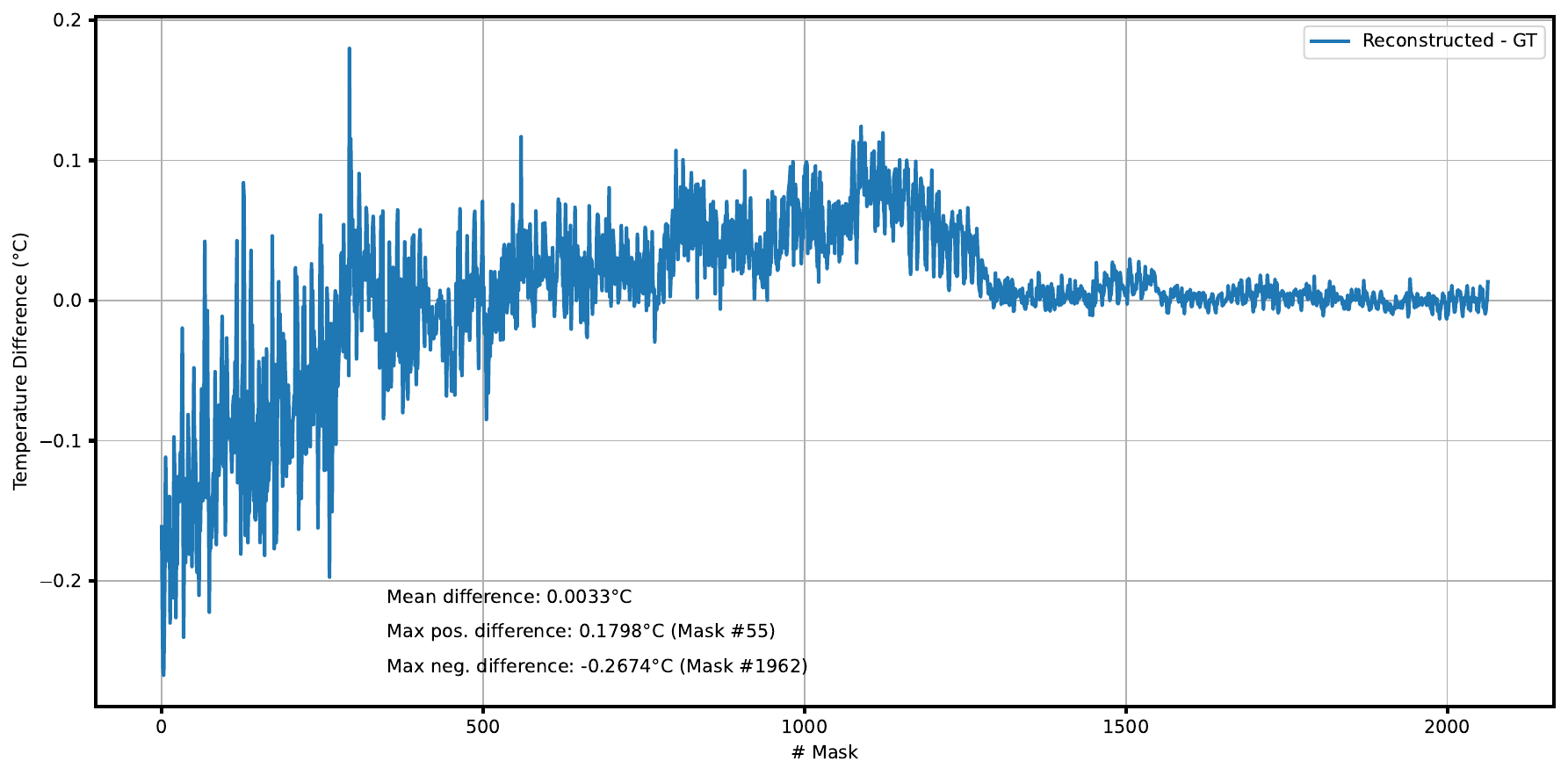}
\caption{\textbf{Difference in GMT between HadCRUT5 temperature field for January 2021 and reconstruction.} Temperature difference between ground truth GMT for January 2021 derived from the HadCRUT5 data set and the reconstructed GMT. We apply all HadCRUT4 derived masks and reconstruct the artificially masked HadCRUT5 temperature field. }
\label{fig:hc5_2052_ts}
\end{figure}

\begin{figure}[b]
\includegraphics[width=\textwidth]{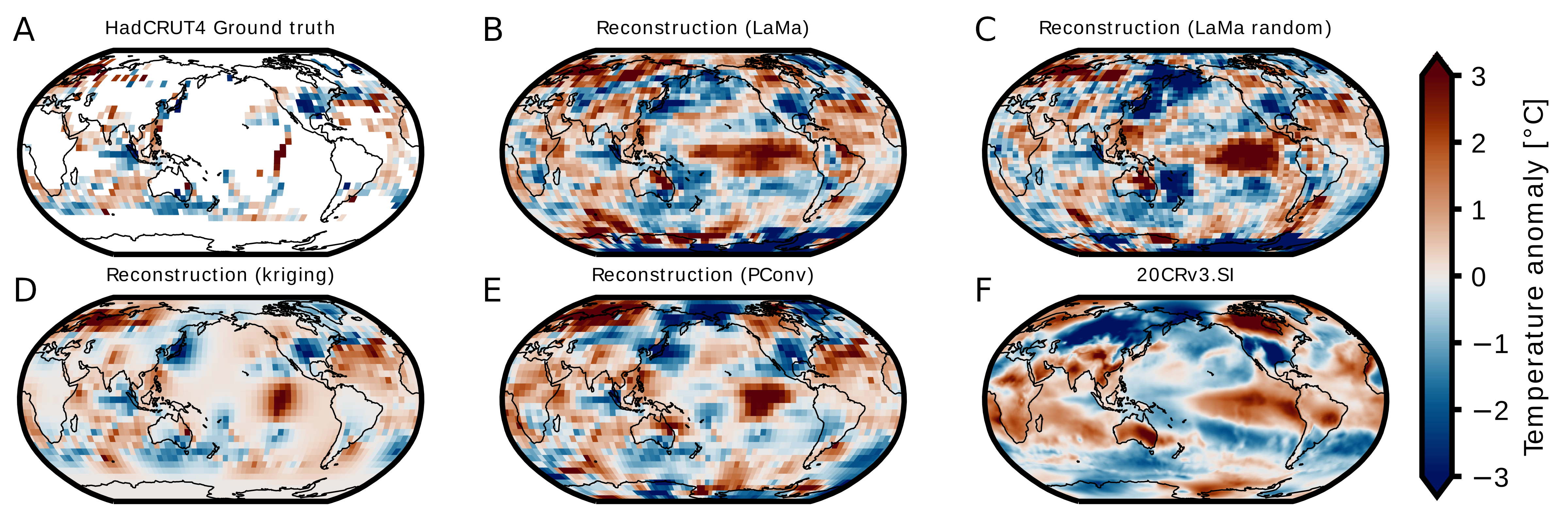}
\caption{\textbf{Reconstructed HadCRUT4 temperature field for November 1877.} \textbf{(A)} Original HadCRUT4 temperature anomaly records. This also corresponds to the input into the trained model. \textbf{(B)} Reconstructed temperature anomalies with LaMa. The strong El Niño is clearly visible in the Pacific. \textbf{(C)} Same but for LaMa random. The spatial extent of the El Niño is clearly visible. \textbf{(D)} Reconstructed temperatures via PConv \cite{kadow_artificial_2020}. \textbf{(E)} Reconstructed temperatures via kriging \cite{cowtan_coverage_2014}. The method fails to reconstruct the spatial extent of the El Niño. \textbf{(F)} Temperature anomalies taken from 20CRv3.SI reanalysis. \cite{compo_twentieth_2011}}
\label{fig:november_1877_hadcrut4}
\end{figure}
\clearpage

\begin{figure}[t]
\includegraphics[width=\textwidth]{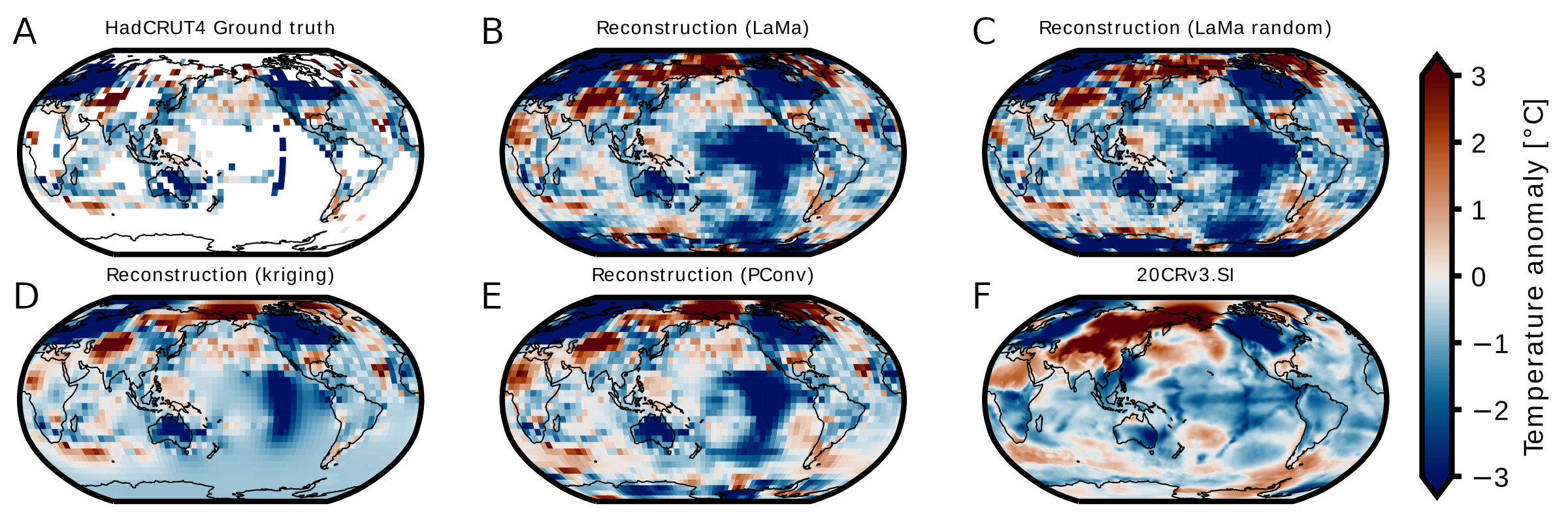}
\caption{\textbf{Reconstructed HadCRUT4 temperature fields for February 1917.} \textbf{(A)} Original HadCRUT4 temperature anomaly records. This also corresponds to the input into the trained model. \textbf{(B)} Reconstructed temperature anomalies with LaMa. The strong La Niña is clearly visible in the Pacific. \textbf{(C)} Same but for LaMa random. The spatial extent of the La Niña is clearly visible. \textbf{(D)} Reconstructed temperatures via PConv \cite{kadow_artificial_2020}. \textbf{(E)} Reconstructed temperatures via kriging \cite{cowtan_coverage_2014}. The method fails to reconstruct the spatial extent of the La Niña. \textbf{(F)} Temperature anomalies taken from 20CRv3.SI reanalysis \cite{compo_twentieth_2011}.}
\label{fig:february_1917_hadcrut4}
\end{figure}
\clearpage

\begin{figure}[t]
\includegraphics[width=\textwidth]{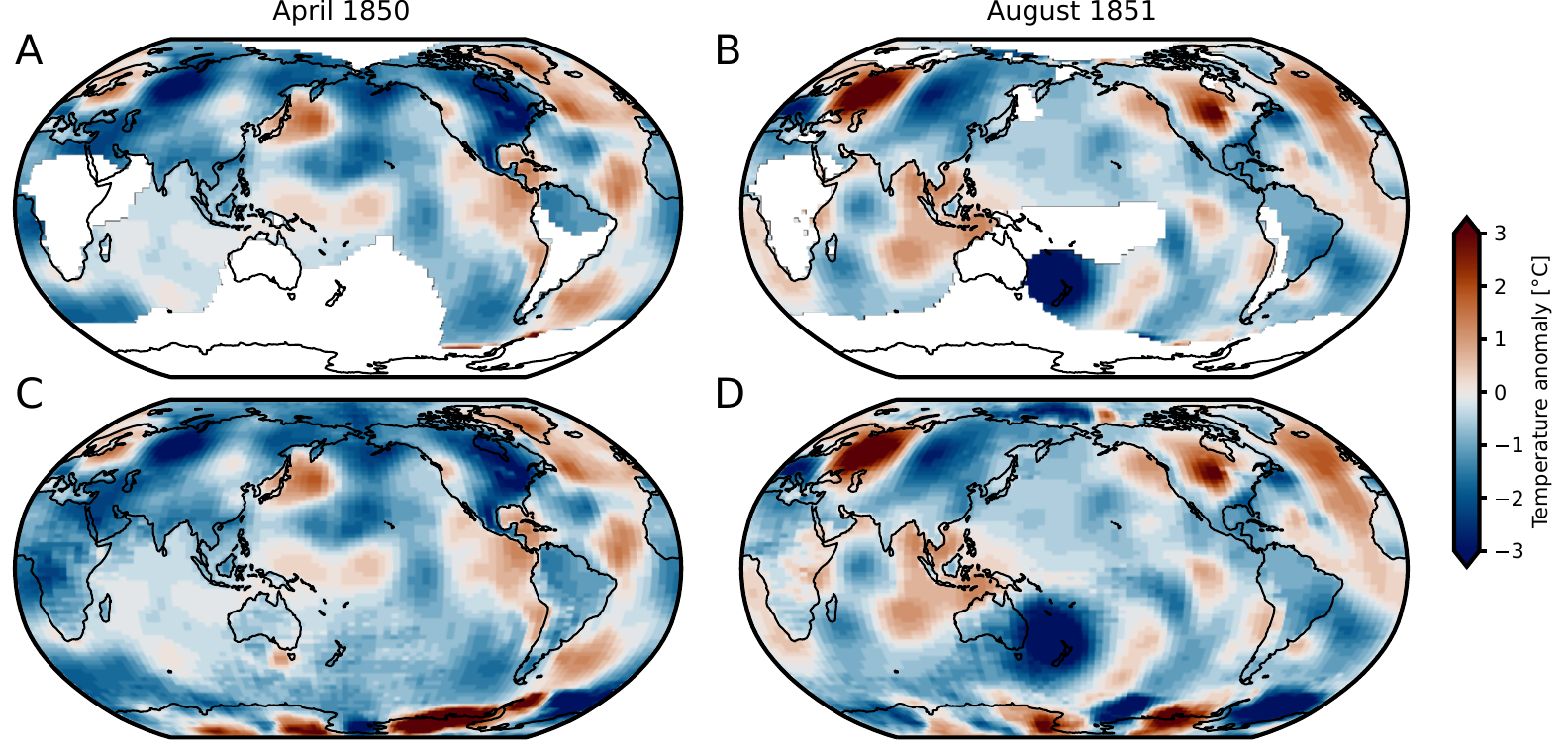}
\caption{\textbf{Reconstruction of two example months for higher resolution BEST data set.} \textbf{(A)} Not inpainted temperature records for April 1850 (BEST dataset,\cite{rohde_berkeley_2020}). \textbf{(B)} Same as \textbf{A} but for August 1851.
\textbf{(C)} Reconstruction of BEST (90x90~px resolution) for April 1850 using LaMa random trained on CMIP5 (72x72~px resolution). \textbf{(D)} Same as \textbf{(C)} but for August 1851. }
\label{fig:best_maps}
\end{figure}

\begin{figure}[t]
\includegraphics[width=\textwidth]{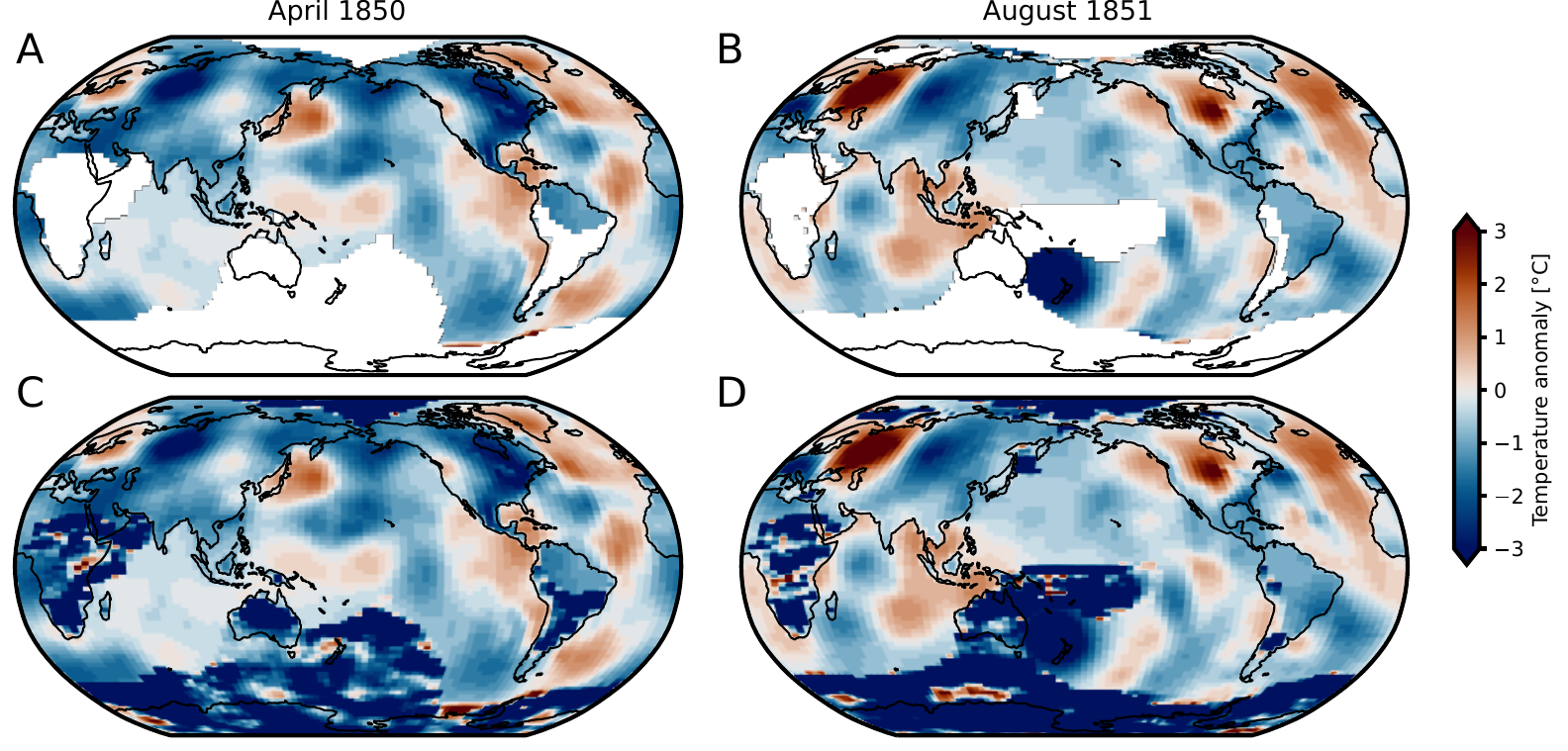}
\caption{\textbf{Example of artifacts of reconstruction for higher resolution BEST data set using LaMa.} \textbf{(A)} Not inpainted temperature records for April 1850 (BEST dataset,\cite{rohde_berkeley_2020}). \textbf{(B)} Same as \textbf{A} but for August 1851.
\textbf{(C)} Reconstruction of BEST (90x90~px resolution) for April 1850 using LaMa trained on CMIP5 (72x72~px resolution). Unseen masks lead to obvious artifacts when filled in with LaMa. \textbf{(D)} Same as \textbf{(C)} but for August 1851.}
\label{fig:best_maps_artifcats}
\end{figure}

\begin{figure}[t]
\includegraphics[width=\textwidth]{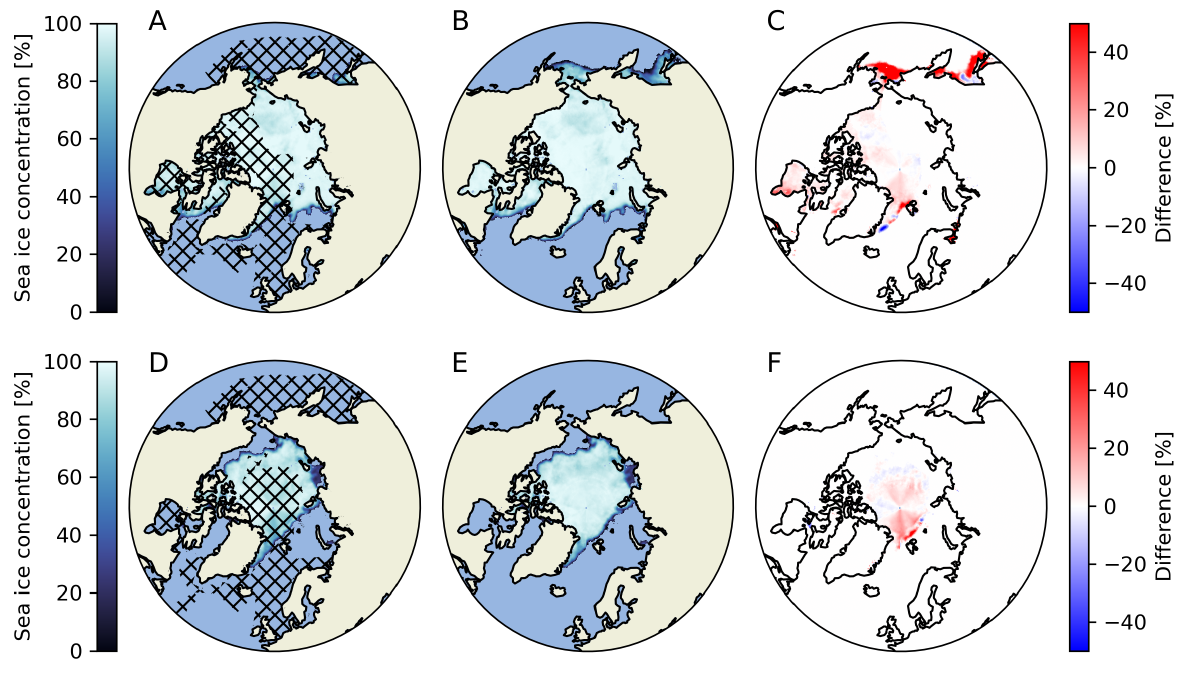}
\caption{\textbf{Example reconstruction of sea ice concentration for two different months in ERA5.} \textbf{(A)} Ground truth of sea ice concentration for held-out December 14, 1979 taken from ERA5 reanalysis \cite{era5_sea_ice}. Hatched area denote regions that are masked for reconstruction. We use LaMa fixed trained on daily sea ice concentration from 1979 to 2022 taken from ERA5 for the reconstruction. \textbf{(B)} Reconstructed sea ice concentration via LaMa fixed. The model is able to reconstruct the spatial extent and concentration of the sea ice reasonably well. \textbf{(C)} Absolute difference between the ground truth and the reconstructed sea ice concentration. Red areas denote overestimated sea ice concentration by the reconstruction, while blue regions denote underestimated sea ice concentration. \textbf{(D,E,F)} Same as \textbf{A}, \textbf{B}, \textbf{C} but for September 16, 1979, respectively.  }
\label{fig:sea_ice}
\end{figure}

\begin{figure}[t]
\includegraphics[width=\textwidth]{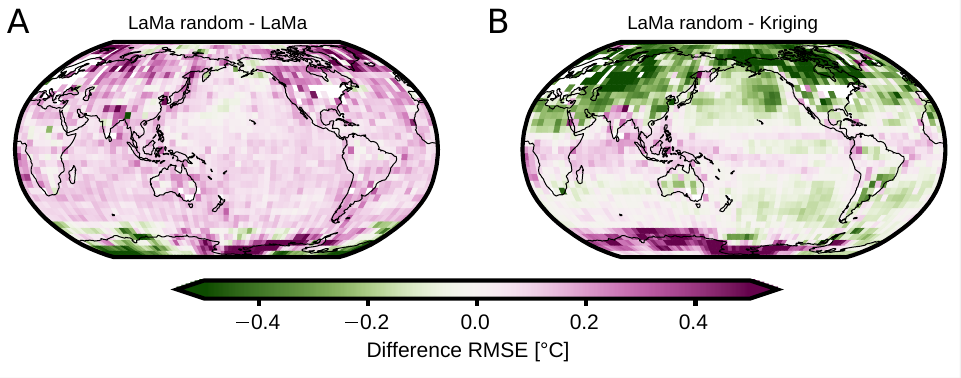}
\caption{\textbf{Difference in site-wise RMSE between LaMa random and LaMa/kriging.} \textbf{(A)} Difference between temporally averaged RMSE at each site between LaMa random and LaMa for held-out CMIP5 member. The white areas denote the regions with available temperature records for the whole time span 1870-2005~AD. Purple areas denote regions where the RMSE of LaMa random is greater than of LaMa. Green areas denote where the RMSE of LaMa random is smaller than for LaMa. LaMa shows a lower RMSE than LaMa random in 82\% of the grid cells. \textbf{(B)} Same as \textbf{A} but for the difference between LaMa random and kriging. LaMa random shows a lower site-wise RMSE than kriging in 65\% of the grid cells (green areas).  }
\label{fig:site_wise_difference_lama}
\end{figure}

\begin{figure}[t]
\includegraphics[width=0.7\textwidth]{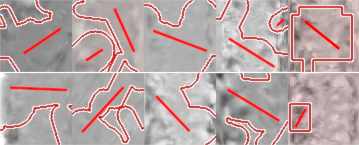}
\caption{\textbf{Example random masks generated during training for LaMa random.} Random masks generated during the training procedure of LaMa random. The masked area is denoted by the red outline and red lines. For details on the generation process see Method section.}
\label{fig:random_masks}
\end{figure}



\end{document}